%% file: PROCI_LaTeX_42.tex
\documentclass[twocolumn,9pt]{article} 

\usepackage[square,numbers,sort&compress,comma]{natbib}

\usepackage{amsmath}
\usepackage{amssymb}
\usepackage{caption}
\usepackage{graphicx}
\usepackage{latexsym}
\usepackage{times}
\usepackage{xcolor}
\definecolor{darkblue}{RGB}{0, 0, 0}   
\definecolor{darkgreen}{RGB}{0, 100, 0} 
\usepackage[pagewise]{lineno}
\usepackage[colorlinks=true,
            linkcolor=darkblue,
            citecolor=blue,
]{hyperref}
\usepackage{cleveref}
\usepackage{subcaption}
\usepackage{cleveref}
\usepackage{pgfplots}

\usepackage{tikz}   
\usepackage{makecell}

\usetikzlibrary{calc,matrix,positioning}
 \usepackage{pgfplots}
 \usepgfplotslibrary{groupplots}
\tikzset{ 
    table/.style={
        matrix of nodes,
        row sep=-\pgflinewidth,
        column sep=-\pgflinewidth,
        nodes={rectangle,text width=1.75em,align=center},
        text depth=0.5ex,
        text height=2.0ex,
        nodes in empty cells,
        align= center,
    },
    activefull/.style={draw=none,fill=blue!60!white},
    activepart/.style={draw=none,fill=blue!20!white},
    show bounding box/.style={
  execute at end picture={
    \path[draw=blue,dashed]
      (current bounding box.south west) rectangle (current bounding box.north east);}}
}
\usepackage{pgfplotstable}
\newcommand{\storelegendimage}[2]{%
  \pgfplotsinvokeforeach{#2}{%
    \expandafter\gdef\csname legendimage@#1\endcsname{%
      \tikzset{every mark/.append style={mark size=1.5pt}}%
      \addlegendimage{#2}%
    }%
  }%
}

\input{acronyms}
\topmargin - 12pt 
\renewenvironment{abstract}%
              {
               \small
               {\bfseries \abstractname}
               \par
               \vspace{10pt}
              }

\renewcommand\abstractname{Abstract}

\newcommand{\nomenclature}
              [1]
              {
               \bgroup
               \flushleft
               \small\bf
               #1
               \par
               \egroup
              }

\renewcommand{\section}
              [1]
              {
               \bgroup
               \flushleft
               \small\bf
               \refstepcounter{section}
               \arabic{section}. #1
               \par
               \egroup
              }

\renewcommand{\subsection}
              [1]
              {
               \bgroup
               \flushleft
               \small\em
               \refstepcounter{subsection}
               \arabic{section}.
               \arabic{subsection}. #1
               \par
               \egroup
              }

\renewcommand{\subsubsection}
              [1]
              {
               \bgroup
               \flushleft
               \small\em
               \refstepcounter{subsubsection}
               \arabic{section}.
               \arabic{subsection}.
               \arabic{subsubsection}. #1
               \par
               \egroup
              }

  \newcommand{\acknowledgement}
              [1]
              {
               \bgroup
               \flushleft
               \small\bf
               #1
               \par
               \egroup
              }

  \newcommand{\sectionbib}
              [1]
              {
               \bgroup
               \flushleft
               \small\bf
               #1
               \par
               \egroup
              }

\begin{document}



\small
\baselineskip 10pt

\setcounter{page}{1}
\title{\LARGE \bf Eigenvalue-based Linear Stability Analysis of Intrinsic Instabilities in Laminar Flames}

\author{{\large Thomas Ludwig Kaiser $^{a,*}$, Peter Munch $^{b}$, Sandra May $^{b}$, Thorsten Zirwes $^{c}$},\\[10pt]
        {\footnotesize \em $^a$ Institute of Fluid Dynamics and Technical Acoustics (ISTA), TU Berlin, Müller-Breslau-Straße 8, Berlin 10623, Germany}\\[-5pt]
        {\footnotesize \em $^b$ Institute of Mathematics, TU Berlin,
        Stra{\ss}e des 17. Juni 136, 10623 Berlin, Germany}\\[-5pt]
        {\footnotesize \em $^c$ Institute for Reactive Flows (IRST), University of Stuttgart, Pfaffenwaldring 31, Stuttgart, 70569, Germany}}

\date{}  

\twocolumn[\begin{@twocolumnfalse}
\maketitle
\rule{\textwidth}{0.5pt}
\vspace{-5pt}

\begin{abstract} 
Intrinsic instabilities of laminar premixed flames play an important role in the dynamics of hydrogen combustion and in the development of predictive models for reacting flows. However, determining their dispersion relations typically relies either on simplified analytical descriptions of the flame front or on computationally expensive direct numerical simulations (DNS). This work develops a generalized eigenvalue problem–based linear stability analysis (GEVP-LSA) framework that predicts the growth rates and spatial structure of intrinsic flame instabilities directly from the linearized governing equations of a 1D base flame. The approach is first validated using the classical Darrieus–Landau configuration, where the numerical results reproduce the analytical dispersion relation and eigenmode structure. The framework is then applied to a model flame of finite thickness governed by the reactive Navier–Stokes equations. The resulting dispersion relations and perturbation fields show excellent agreement with corresponding DNS results while reducing the computational effort by a factor of $10^8$. The proposed method therefore provides an efficient and accurate tool for studying intrinsic flame instabilities and offers a scalable foundation for future stability analyses of more complex reacting-flow configurations relevant to combustion modeling and large-eddy simulations.
\end{abstract}

\vspace{10pt}

{\bf Novelty and significance statement}

\vspace{10pt}

Existing approaches for determining dispersion relations (DRs) of intrinsic flame instabilities fall into two categories. Analytical approaches provide valuable physical insight but rely on simplifying assumptions that often lead to quantitatively inaccurate predictions. Alternatively, Direct Numerical Simulations (DNS) can determine DRs with high accuracy; however, their high computational cost severely limits exploration of large parameter spaces.

This work establishes a third approach in which the stability problem is formulated directly as a generalized eigenvalue problem derived from the linearized 1D reactive Navier–Stokes equations. This formulation enables the direct computation of instability growth rates and eigenmodes while retaining the full governing physics and reducing computational cost by up to eight orders of magnitude compared to DNS. The proposed framework therefore enables systematic investigations of intrinsic flame instabilities and lays the groundwork for studying phenomena such as thermodiffusive instabilities with DNS-level accuracy but at a fraction of the computational cost.

\vspace{5pt}
\parbox{1.0\textwidth}{\footnotesize {\em Keywords:} Intrinsic flame instability; Linear stability analysis; Direct Numerical Simulation}
\rule{\textwidth}{0.5pt}
*Corresponding author.
\vspace{5pt}
\end{@twocolumnfalse}] 

\section{Introduction\label{sec:introduction}} \addvspace{10pt}

The energy transition is driving the demand for new combustor technologies for hydrogen–air burners, due to the flashback tendencies of the novel fuel and unconventional thermoacoustic behavior.
The current lack of reliable \gls{LES} sub-grid models for intrinsic instabilities in hydrogen flames hinders the development of these new combustor technologies. The main reason for the insufficient modeling capabilities is the complex thermodiffusive mechanisms and the Soret effect that dominate lean hydrogen flames. Understanding these effects is essential for developing accurate sub-grid models.

Analytical models, such as those based on the work of Markstein~\cite{Markstein1951}, Matalon et al.~\cite{Matalon1982,Matalon2003,matalon2007intrinsic}, Sivashinsky et al.~\cite{Sivashinsky1977,Frankel1982}, and others~\cite{creta2020propagation, Joulin1979,Frankel1982,Pelce1982,Sharpe2003, Clavin1982}, provide valuable insight into the underlying driving mechanisms but are not quantitatively predictive for all flame configurations~\cite{lapenna2023hydrogen,pitsch2024transition}. In recent years, a significant number of \gls{DNS} studies have focused on this topic, substantially advancing the understanding of intrinsic instabilities in premixed flames---particularly for lean hydrogen mixtures, where thermodiffusive mechanisms and the Soret effect strongly influence both growth and flame morphology~\cite{howarth2023thermodiffusively}. \gls{DNS} investigations have documented synergistic interactions~\cite{berger2022synergistic}, pressure and curvature sensitivities~\cite{attili2021effect}, and pronounced Soret-driven stretch responses~\cite{zirwes2025assessment,zirwes2024role}. 
Typically, the \gls{DNS} methodology involves perturbing a laminar flame front in a 2D or 3D configuration and determining the \gls{DR} by measuring the temporal growth rate of the perturbations as a function of wavenumber. The parameter space to be explored with this approach is, however, far too large to be efficiently covered, as detailed diffusion and kinetics models are required. The recent work of Al Kassar et al.~\cite{al2024efficient}, who perturbed a laminar flame front simultaneously with several wavenumbers and determined their independent growth in post-processing, highlights the urgent need for more efficient methods.

A methodology that has been applied to study instability mechanisms across many fields of physics is eigenvalue-based/operator-based \gls{LSA}. The groundwork for this analysis was laid by Helmholtz~\cite{Helmholtz_1868}, Rayleigh~\cite{rayleigh_instability_1878} and Thomson~\cite{Thomson1871} who used perturbation approaches to study the stability of shear flows. Eigenvalue-based stability analysis of shear flows was introduced later by Orr~\cite{Orr1907} and Sommerfeld~\cite{Sommerfeld1909}. Since then, the eigenvalue-based \gls{LSA}—both analytical and numerical—has become state-of-the-art for investigating stability problems in a significant number of scientific fields. Examples include the stability analysis of the Lotka-Volterra equations, which model the dynamics of predator-prey populations in zoology~\cite{hofbauer1998}, magnetohydrodynamics~\cite{chandrasekhar1961hydrodynamic}, and astrophysics~\cite{Balbus1991}. Particularly modern numerical discretization techniques allow the use of eigenvalue-based \gls{LSA} without making strong simplifying assumptions. Such numerical operator-based linear analysis has been applied in recent years to both laminar~\cite{blanchard2015response,wang2022linear,avdonin2019thermoacoustic,meindl2021spurious} and turbulent~\cite{manoharan2015absolute,kaiser2019prediction,casel2022resolvent,kaiser2023modelling,chauhan_sotic_2025} flame configurations to study thermoacoustics and turbulence flame interaction. However, in the context of the analysis of intrinsic cellular instabilities in laminar flames, the application of such approaches is still underdeveloped. 

The present work addresses this research gap by pursuing an eigenvalue-based \gls{LSA} approach that directly targets the \gls{DR} through a \gls{GEVP} formulated on a 1D base flame. The objective is to deliver \gls{DR} predictions of comparable quality to \gls{DNS}, but at only a fraction of the computational cost. First, the \gls{GEVP}-based \gls{LSA} will be performed for the classical Darrieus--Landau flame configuration, and subsequently on a solution to the full reacting Navier--Stokes equations. 

The manuscript is structured as follows. Section~\ref{LSAMethod} introduces the eigenvalue-based \gls{LSA}. The numerical approach will then be tested to reproduce the analytical solution to the classical Darrieus-Landau flame configuration in Section~\ref{sec:Landau}. The methodology will then be extended to a flame of finite thickness governed by the reacting Navier-Stokes equations in Section~\ref{sec:finite}, where the results of the \gls{LSA} will be validated against \gls{DNS} data. Finally, the study is concluded in Section~\ref{sec:conclusion}.

\section{Theory of eigenvalue-based linear stability analysis\label{LSAMethod}}\addvspace{10pt}

Consider a set of closed transport equations governing the state variables $\mathbf{\Phi} (x,y,t)$ of a 3D laminar planar flame, defined by the non-linear operator $\mathcal{N}$, reading
\begin{equation}
    \frac{\partial \mathbf{\Phi} (x,y,z,t)}{\partial t} = \mathcal{N}(\mathbf{\Phi} (x,y,z,t)).\label{eq:transportEq}
\end{equation}
To obtain the linearized governing equations, the state variables ($\mathbf{\Phi} (x,y,z,t)$) are decomposed in a temporal mean part, indicated by an overline, and a small ($\epsilon \ll 1$) fluctuation, indicated by a tilde:
\begin{equation}
    \mathbf{\Phi}(x,y,z,t) = \overline{\mathbf{\Phi}}(x) + \epsilon \widetilde{\mathbf{\Phi}}(x,y,z,t).\label{eq:doubleDecomposition}
\end{equation}
To investigate the stability of the laminar planar flame oriented perpendicular to the $x$-axis, it is convenient to express the fluctuating part as a Fourier mode and its complex conjugate~($\text{c.c.}$).
\begin{equation}
    \widetilde{\mathbf{\Phi}}(x,y,t) = \widehat{\mathbf{\Phi}} (x) \text{exp}\left( - i \omega t + i k_y y + i k_z z\right) + \text{c.c.},\label{eq:FourierMode}
\end{equation}
where $\omega$ is the angular frequency and $k_y$ and $k_z$ are the wave numbers in $y$ and $z$-direction, respectively. 
In general, both $\omega$ and $k$ are complex quantities. The real part of the angular frequency $\omega_\text{r}$ describes the harmonic frequency of the oscillation in time, and its imaginary part $\omega_\text{i}$ the temporal growth rate. Analogously, the real and imaginary parts of $k_y$ and $k_z$ are the wavenumbers and growth rates in the respective directions. For intrinsic instabilities in laminar flames, we expect, however, that $k_y$, $k_z$, and $\omega$ are not complex. Unlike most instability mechanisms, such as cylinder wake modes \cite{barkley2006linear}, intrinsic flame instabilities grow, but do not oscillate in time. Hence, for an unstable laminar flame, we expect $\omega_r = 0$ and $\omega_i>0$. At the same time, in the present configuration, the base state has no gradients in $y$ and $z$, so any \gls{LSA} has translational symmetry along the flame front. A non-zero imaginary part of $k_y$ or $k_z$ would imply exponential growth or decay along that direction, which would break that symmetry. Hence, physical modes for a planar flame must have real values for $k_y$ and $k_z$ only. 

Inserting Eqs.~(\ref{eq:doubleDecomposition}--\ref{eq:FourierMode}) into the transport equations (Eq.~(\ref{eq:transportEq})) allows to derive the linearized governing equations. These take the form of a \gls{GEVP}:
\begin{equation}
\omega \mathbf{B} \widehat{\mathbf{\Phi}}(x) =  \mathbf{A}(\overline{\mathbf{\Phi}}(x),k_y,k_z) \widehat{\mathbf{\Phi}}(x). \label{eq:GEVP}
\end{equation}
Equation~(\ref{eq:GEVP}) yields the \gls{DR} of the laminar flame, since it links the wave number $k_y$ to the angular frequency, $\omega$, where $\omega_i$ is the temporal growth rate. The equation also illustrates the advantages of the approach in determining the \gls{DR} in comparison to \gls{DNS}. While in the \gls{DNS} the non-linear governing equations are integrated in all spatial directions and time, the linearization of the equations yields a system where the discretization in time as well as in $y$ and $z$ is obtained by the Fourier decomposition. Therefore, it is sufficient to numerically discretize the \gls{GEVP} in 1D, i.e., in $x$-direction. It can be expected that this is significantly cheaper than time-stepping the set of non-linear equations in 3D and determining the \gls{DR} a posteriori. The associated disadvantage that complex-valued quantities have to be handled is minor compared to the efficiency gains.
\section{Stability of an infinitely thin flame front\label{sec:Landau}}
\subsection{Linear governing equations of the infinitely thin flame front}\label{sec:Theory:Landau}\addvspace{10pt}

\begin{figure}
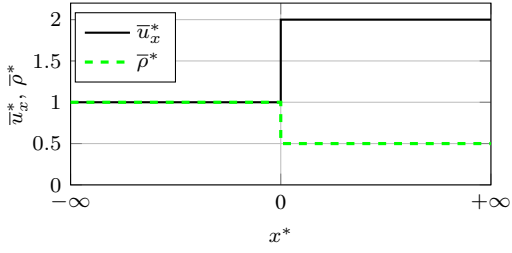

    \centering
    \include{figures/LandaMean}\vspace{-0.3cm}
    \vspace{-0.5cm}
    \caption{Unperturbed base states of the Darrieus--Landau flame configuration for $\overline{u}^*_{x,\text{u}}=\overline{\rho}_\text{u}^*=1$ and $\sigma = 2$.}
    \label{fig:Landau_mean}
\end{figure}
The simple, inviscid flame model discussed by Darrieus~\cite{darrieus1938propagation} and Landau~\cite{Landau1944} assumes an infinitely thin flame, leading to a jump in density, $\rho$, and velocity, $u$, of strength $\sigma = \frac{\rho_1}{\rho_2} = \frac{u_2}{u_1}$, where the indices indicate the upstream and downstream gas state, illustrated in Fig.~\ref{fig:Landau_mean}. Across this flame front, conservation equations of mass and momentum in the absence of body and viscous forces hold. For a 2D flame, where $k_z=0$ and $\widehat{u}_z=0$, their linearization reads
\begin{equation}
    \frac{\partial \widehat{u}^*_x}{\partial x^*} + i k^*_y \widehat{u}^*_y  =0, \label{eq:conti}
\end{equation}
\begin{equation}
    \omega^* \widehat{u}^*_x
    + \overline{u}^*_x \frac{\partial \widehat{u}^*_x}{\partial x^*} 
    = - \frac{1}{\rho^*} \frac{\partial \widehat{p}^*}{\partial x^*}, \label{eq:momentumx}
\end{equation}
\begin{equation}
    \omega^* \widehat{u}^*_y 
    + \overline{u}^*_x \frac{\partial \widehat{u}^*_y}{\partial x^*} 
    = - \frac{1}{\rho^*} i k^*_y \widehat{p^*}.
\end{equation}

where $*$ indicates a non-dimensional quantity. 
At the density and velocity jump, Landau gives the following contact conditions:
\begin{equation}
    \widehat{p}^*_1 = \widehat{p}^*_2,
\end{equation}
\begin{equation}
    \widehat{u}^*_{x,1} =\widehat{u}^*_{x,2},
\end{equation}
\begin{equation}
\omega^* \widehat{u}^*_{1,y}+ k^*_y \overline{u}^*_{1,x}  \widehat{u}^*_{1,x} = \omega^* \widehat{u}^*_{2,y}+ k^*_y \overline{u}^*_{2,x}  \widehat{u}^*_{2,x}.\label{eq:BCuy}
\end{equation}
The set of Equations ~(\ref{eq:momentumx}-\ref{eq:BCuy}) is discretized using FEniCS~\cite{baratta_dolfinx_nodate}.
Eq.~(\ref{eq:BCuy}) is a jump condition introducing a discontinuity in $\widehat{u}^*_{1,y}$. As a consequence, the equations are discretized using a \gls{DG} approach. 
Quadratic polynomials are used for the velocities and linear elements for pressure. The advection term is stabilized using upwinding. The 1D domain is discretized in the interval between $x^*=-50$ to $x^*=50$ with 60 elements, where the flame is located at $x^*=0$. A constant factor mesh grading is introduced with the finest element size of $\Delta x^*=0.04$ located at the flame front. Consistent with the analytical solution obtained by Landau, indicating an exponential decay of all fluctuations in both negative and positive $x$-direction, homogeneous Dirichlet boundary conditions are applied at both the upstream and downstream domain boundary. The relatively small number of degrees of freedom allows to solve the discretized \gls{GEVP} for all its eigenvalues using LAPACK~\cite{lapack99}.

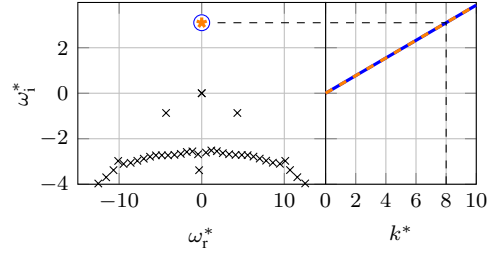
\begin{figure}
    \centering
    \input{figures/spectrumLandau}
    \caption{Stability results for Landau's flame configuration for $\overline{u}^*_{x,\text{u}}=\overline{\rho}_\text{u}^*=1$ and $\sigma=2$; left: stability spectrum with analytical solution (\ref{line:analyticSpectrum}), as well as physical (\ref{line:numSpectrumPhysical}) and spurious (\ref{line:numSpectrumSpurious}) numerical solutions for $k^*_y=8$; right: analytical \gls{DR} based on Landau~\cite{Landau1944} (\ref{line:analytical}) and numerical \gls{DR} (\ref{line:numerical}).}
    \label{fig:LandauSpectrum}
\end{figure}
\subsection{Numerical results for the stability of the infinitely thin flame front}\addvspace{10pt}
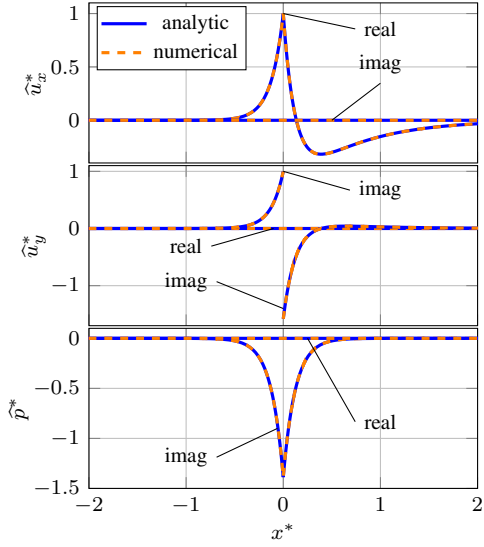
\begin{figure}[t]
    \centering
\input{figures/LandauModes}
\caption{Solutions to the Darrieus--Landau flame configuration: Velocity and pressure perturbations.}
    \label{fig:LanauModes}
\end{figure}
Figure~\ref{fig:LandauSpectrum} compares the spectrum obtained by \gls{GEVP} \gls{LSA} to the analytical solution. The agreement between the numerical and analytical \glspl{DR} is excellent across the full range of wavenumbers within the unstable regime. 
The eigenmode structures depicted in Fig.~\ref{fig:LanauModes} further substantiate this assessment. The velocity and pressure perturbations exhibit the characteristic exponential decay away from the flame front predicted by the analytical solution, with the expected phase relationship between the streamwise and transverse velocity components. 
The quantitative match with Landau’s analytical solution demonstrates that the \gls{GEVP} formulation captures both the temporal amplification behavior and the spatial modal structure of intrinsic laminar flame instabilities. Consequently, this benchmark validates the applicability of the proposed \gls{GEVP}-based \gls{LSA} approach as a reliable foundation for the forthcoming analysis of a more complex and realistic set of governing equations.

\section{Stability of a model flame with finite thickness\label{sec:finite}}
\subsection{Direct Numerical Simulations \label{subsec:dns}} \addvspace{10pt}

To provide reference solutions for the instability growth rate in the reacting Navier--Stokes equations, \gls{DNS} are performed for a simplified flame configuration by solving the following equations with an in-house OpenFOAM-based code~\cite{zirwes2023assessment}:
\begin{equation}
\frac{\partial \rho}{\partial t} + \nabla \cdot (\rho \mathbf{u}) = 0, \label{eq:NonLinear:Conti}
\end{equation}
\begin{equation}
\rho\Bigl(\frac{\partial \mathbf{u}}{\partial t} + (\mathbf{u}\!\cdot\!\nabla)\mathbf{u}\Bigr)
= -\,\nabla p + \nabla\!\cdot\!\boldsymbol{\tau},\label{eq:NonLinear:Momentum}
\qquad
\end{equation}
\begin{equation}
\boldsymbol{\tau} = \mu \left( \nabla \mathbf{u} + (\nabla \mathbf{u})^{T}  
- \frac{2}{3}\nabla \cdot (\mathbf{u})\mathbf{I}\right) ,\label{eq:NonLinear:tau}
\end{equation}
\begin{equation}
\rho \frac{\partial c}{\partial t} + \rho \mathbf{u}\cdot\nabla c
= \nabla \cdot \bigl( \Gamma \nabla c\bigr)\label{eq:NonLinear:c}
+ \dot{\Omega},
\end{equation}
where $c$ is the reaction progress, $\Gamma=\mu=5\times10^{-5}$\,m$^2$/s, implying unity Lewis, Schmidt and Prandtl numbers ($Le=Sc=Pr=1$), and the source term $\dot{\Omega}$ modeling the reaction equals
\begin{equation}
    \dot{\Omega} = A (1-c) \text{exp} \left( - \frac{T_\text{a}}{T}\right).\label{eq:NonLinear:sourceTermn}
\end{equation}

In this formulation based on an Arrhenius reaction, the term $(1-c)$ models the depletion of the deficient species, while the temperature necessary to initiate the reaction is modeled by an activation temperature of $T_\text{a}=15,000\,\text{K}$. The pre-exponential factor takes the value $A= 10^{8}\, \text{kg}/\text{m}^{3}/\text{s}$. The temperature is related to the progress variable by the linear relation
\begin{equation}
    T=T_u+(T_b-T_u)c,\label{eq:NonLinear:Temperature}
\end{equation}
where $T_u=300$\,K and $T_b=2400$\,K correspond to the unburnt and burnt gas temperatures, respectively. Density is calculated from
\begin{equation}
    \rho=\frac{p_0}{R_sT}\label{eq:nonLinear:perfectGasLaw}
\end{equation}
with $p_0=10^5$\,Pa and the specific gas constant $R_s=(\mathcal{R}/M)$, where $\mathcal{R}$ is the universal gas constant and $M$ the constant molecular weight $M=22$\,kg/kmol. With these settings, which mimic a generic hydrogen flame, the properties of the freely propagating flame were calculated from a 1D flame solution implemented in Cantera, yielding the unperturbed flame speed $s_L^0=0.766$\,m/s and thermal flame thickness~$\delta_{\rm th}=0.128$\,mm.



\begin{figure}[t]
    \centering
    \includegraphics[width=0.95\linewidth]{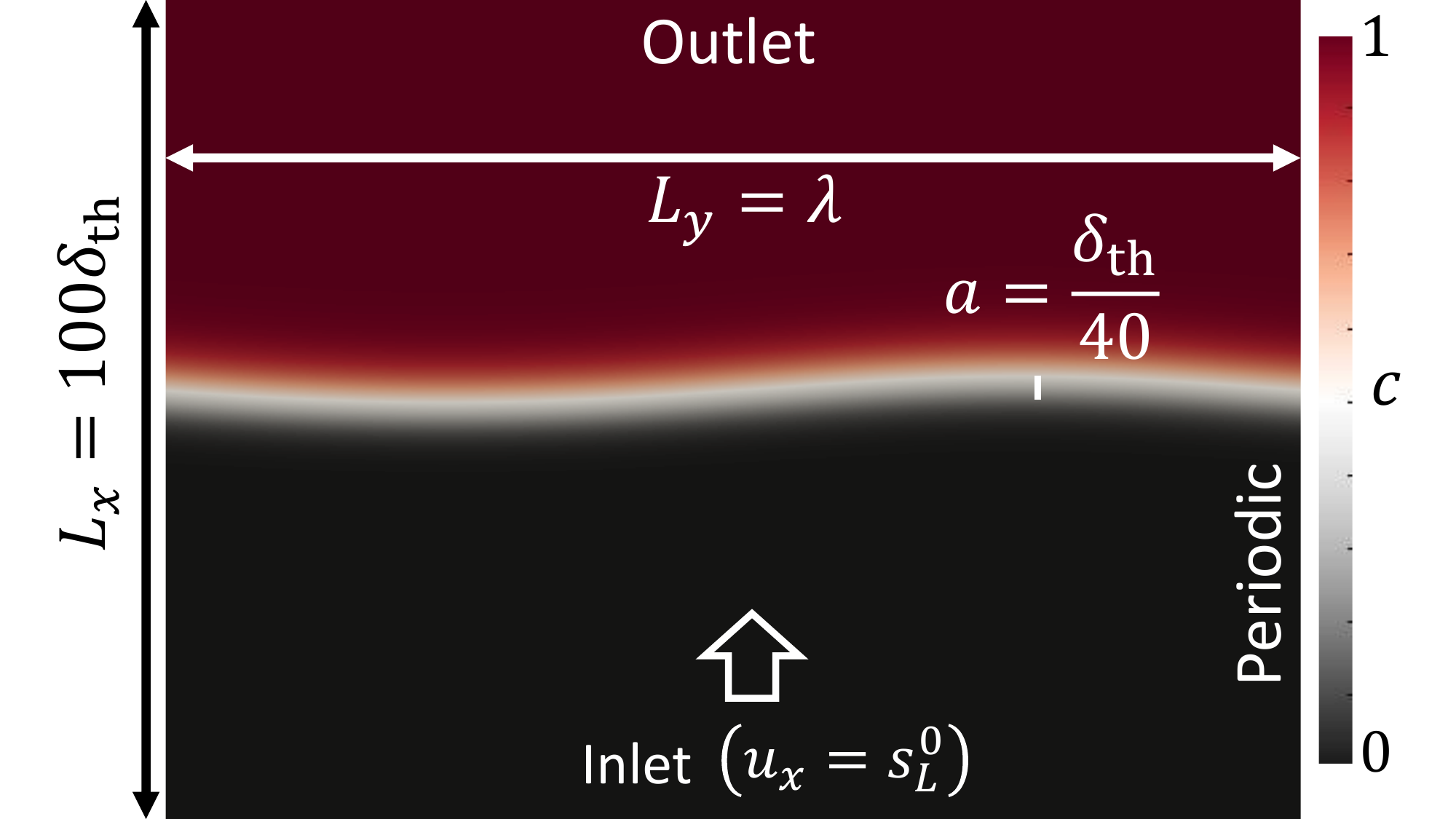}%
    \vspace*{4pt}\caption{\footnotesize Schematic of the 2D \gls{DNS} setup (not to scale).}
    \label{fig:setup}
\end{figure}

To determine instability growth rates, a 2D computational domain  (Fig.~\ref{fig:setup}) with dimensions $L_x=100\delta_{\rm th}$ and $L_y=\lambda$ is employed. The flame is initialized in the center of the domain from a 1D solution of the reference flame. This 1D profile is then perturbed with a sinusoidal perturbation with wavelength~$\lambda=2\pi/k$ and amplitude $a=\delta_{\rm th}/40$. Mesh resolution is equidistant with $\Delta x=\Delta y=\delta_{\rm th}/20$. The lateral sides are periodic boundaries. An inlet provides fresh gases with $c=0$ and a velocity of $u_x=s_L^0$. For the 3D cases, $L_y=L_z=\lambda$ and the dimension in flow direction is reduced to $L_x=40\delta_{\rm th}$ and the flame is perturbed with two sinusoidal functions in the $x$ and $y$ direction with the same wavelength~$\lambda$.

\subsection{Stability of the laminar flame front with finite thickness}\addvspace{10pt}

The linearization of the governing equations in Eqs.~(\ref{eq:NonLinear:Conti}--\ref{eq:NonLinear:c}) follows the same procedure as that used for Eqs.~(\ref{eq:momentumx}--\ref{eq:conti}). Applying this approach yields the following set of linearized equations:
\begin{equation}
\begin{aligned}
i\omega\,\widehat{\rho} &=
\overline{u}_x\,\frac{\mathrm{d}\widehat{\rho}}{\mathrm{d}x} +\widehat{u}_x\,\frac{\mathrm{d}\overline{\rho}}{\mathrm{d}x}
+\widehat{\rho}\,\frac{\mathrm{d}\overline{u}_x}{\mathrm{d}x} \\
&+\overline{\rho}\left(
\frac{\mathrm{d}\widehat{u}_x}{\mathrm{d}x}
+ i k_y \widehat{u}_y
+ i k_z \widehat{u}_z
\right) ,\label{eq:lin:conti}
\end{aligned}
\end{equation}

\begin{equation}
\begin{aligned}
i\omega\,\overline{\rho}\,\widehat{u}_x =&
\overline{\rho}\left(
\overline{u}_x\frac{\mathrm{d}\widehat{u}_x}{\mathrm{d}x}
+ \widehat{u}_x\,\frac{d\overline{u}_x}{dx}
\right)
+ \widehat{\rho}\left(\overline{u}_x\,\frac{\mathrm{d}\overline{u}_x}{\mathrm{d}x}\right)\\
+ &\frac{\mathrm{d}\widehat{p}}{\mathrm{d}x} 
\;-\left(
\frac{\mathrm{d}\widehat{\tau}_{xx}}{\mathrm{d}x}
+ i k_y\,\widehat{\tau}_{xy}
+ i k_z\,\widehat{\tau}_{xz}
\right),\label{eq:lin:ux}
\end{aligned}
\end{equation}

\begin{equation}
\begin{aligned}
i\omega\,\overline{\rho}\,\widehat{u}_y &=
\overline{\rho}\left(
\overline{u}_x\frac{\mathrm{d}\widehat{u}_y}{\mathrm{d}x}
\right) 
+ i k_y \widehat{p} \\
&-\left(
\frac{\mathrm{d}\widehat{\tau}_{yx}}{\mathrm{d}x}
+ i k_y\,\widehat{\tau}_{yy}
+ i k_z\,\widehat{\tau}_{yz}
\right),\label{eq:lin:uy}
\end{aligned}
\end{equation}
\begin{equation}
\begin{aligned}
i\omega\,\overline{\rho}\,\widehat{u}_z &=
\overline{\rho}\,
\overline{u}_x\frac{\mathrm{d}\widehat{u}_z}{\mathrm{d}x}
+ i k_z \widehat{p} \\
&-\left(
\frac{\mathrm{d}\widehat{\tau}_{zx}}{\mathrm{d}x}
+ i k_y\,\widehat{\tau}_{zy}
+ i k_z\,\widehat{\tau}_{zz}
\right),\label{eq:lin:uz}
\end{aligned}
\end{equation}
\begin{equation}
\begin{aligned}
i\omega\,\overline{\rho}\,\widehat{c}
&=
\overline{\rho} \,
\overline{u}_x\frac{\mathrm{d}\widehat{c}}{\mathrm{d}x}
+ \overline{\rho} \widehat{u}_x\,\frac{\mathrm{d}\overline{c}}{\mathrm{d}x}
+ \widehat{\rho}\overline{u}_x\,\frac{\mathrm{d}\overline{c}}{\mathrm{d}x} \\
&-\left[
\Gamma\,\frac{\mathrm{d}^2\widehat{c}}{\mathrm{d}x^2}
- (k_y^2+k_z^2)\,\Gamma\,\widehat{c}
\right]
-\widehat{\dot{\Omega}}.\label{eq:lin:c}
\end{aligned}
\end{equation}
Here, Eq.~(\ref{eq:lin:conti}) corresponds to the linearized continuity equation, Eq.~(\ref{eq:lin:ux}--\ref{eq:lin:uz}) to the linearized momentum equations in $x$, $y$, and $z$ directions respectively, while Eq.~(\ref{eq:lin:c}) is the linearized transport equation of the progress variable. The fluctuations of the strain rate tensor are given by the linearization of Eq.~(\ref{eq:NonLinear:tau}), reading
\begin{equation}
\widehat{\tau}_{xx}
=
\frac{4}{3}\mu\,\frac{\mathrm{d}\widehat{u}_x}{\mathrm{d}x}
-\frac{2}{3}\mu\left(
i k_y \widehat{u}_y
+ i k_z \widehat{u}_z
\right),
\end{equation}
\begin{equation}
    \widehat{\tau}_{yy}
=
\frac{4}{3}\mu\, i k_y \widehat{u}_y
-\frac{2}{3}\mu\left(
\frac{\mathrm{d}\widehat{u}_x}{\mathrm{d}x}
+ i k_z \widehat{u}_z
\right),
\end{equation}
\begin{equation}
    \widehat{\tau}_{zz}
=
\frac{4}{3}\mu\, i k_z \widehat{u}_z
-\frac{2}{3}\mu\left(
\frac{\mathrm{d}\widehat{u}_x}{\mathrm{d}x}
+ i k_y \widehat{u}_y
\right),
\end{equation}
\begin{equation}
    \widehat{\tau}_{xy}=\widehat{\tau}_{yx}
=
\mu\left(
\frac{\mathrm{d}\widehat{u}_y}{\mathrm{d}x}
+ i k_y \widehat{u}_x
\right),
\end{equation}
\begin{equation}
    \widehat{\tau}_{xz}=\widehat{\tau}_{zx}
=
\mu\left(
\frac{\mathrm{d}\widehat{u}_z}{\mathrm{d}x}
+ i k_z \widehat{u}_x
\right),
\end{equation}
\begin{equation}
    \widehat{\tau}_{yz}=\widehat{\tau}_{zy}
=
\mu\left(
i k_y \widehat{u}_z
+ i k_z \widehat{u}_y
\right).
\end{equation}

Finally, the set of equations is closed by the linearizations of the source term in the transport equation of the progress variable (see Eq.~(\ref{eq:NonLinear:sourceTermn})), yielding the linear fluctuation in reaction rate
\begin{equation}
\begin{aligned}
        \widehat{\dot{\Omega}} &= A  (1-\overline{c})\left(\frac{T_\text{a}}{\overline{T}^2}  \exp\left(-\frac{T_\text{a}}{\overline{T}}\right)\right) \widehat{T} \\ &-A \,\exp \left(-\frac{T_\text{a}}{\overline{T}}\right) \widehat{c}.
\end{aligned}
\end{equation} 
Similarly, the linearization of Eq.~(\ref{eq:NonLinear:Temperature}), yields the linear fluctuation of temperature
\begin{equation}
    \widehat{T} = (T_\text{b}- T_\text{u}) \widehat{c}.\label{eq:lin_T}
\end{equation}
Unlike in the Darrieus–Landau configuration, all quantities in the non-linear and linear equations are continuous. As a consequence, a \gls{CG}-\gls{FEM} approach is sufficient. The calculations for the model flame with finite thickness are performed using the linear flow solver FELiCS~\cite{kaiser_felics_2023}. Quadratic Taylor-Hood elements are applied, using second order basis functions for all velocity components and the progress variable, and first order basis functions for pressure. 

Fluctuations are expected to follow $\exp(k x)$ upstream of the flame, and $\exp(-k x)$ downstream of the flame~\cite{Landau1944}. To ensure a natural decay towards the boundaries for small wave numbers $k$, a sufficiently large domain is needed.
Therefore, the 1D computational domain spans from $x\approx-10\, \text{m}$ to $x\approx10\, \text{m}$. The flame is located in the center of the domain by enforcing $\overline{c}(x=0)=0.5$. In the vicinity of the flame, i.e. from $x=-5 \cdot 10^{-4}\, \text {m}$ to $x=5 \cdot 10^{-4}\, \text{m}$, the element size is $\Delta x = 2 \cdot 10^{-5}~\text{m}$. Towards the upstream and downstream boundaries, the mesh is coarsened with a constant element-to-element size ratio, of approximately $1.2$, leading to a total number of 182 elements. A mesh convergence study showed that refining the computational grid by four times leads to changes in calculated growth rates of less than one percent.


At both the upstream and downstream boundary, homogeneous Dirichlet boundary conditions are imposed. After the discretization, the system matrices $\mathbf{A}$ and $\mathbf{B}$ (see Eq.~\ref{eq:GEVP}) are of size $1643\times1643$. Solving a \gls{GEVP} of that size for all eigenvalues as in the infinitely thin flame front is not a viable approach. Therefore, the \gls{GEVP} resulting from Eqs.~(\ref{eq:lin:conti})--(\ref{eq:lin:c}) is solved using the shift and invert methodology implemented in the SLEPc~\cite{hernandez_slepc_2005} library.

\subsection{Stability of the two-dimensional model flame with finite thickness\label{sec:modelFlame2D}}
\addvspace{10pt}
\begin{figure}[t]
    \centering
\input{figures/baseState}
\caption{Unperturbed 1D base states resulting from Eqs.~(\ref{eq:NonLinear:Conti}--\ref{eq:nonLinear:perfectGasLaw}).}
    \label{fig:baseState}
\end{figure}
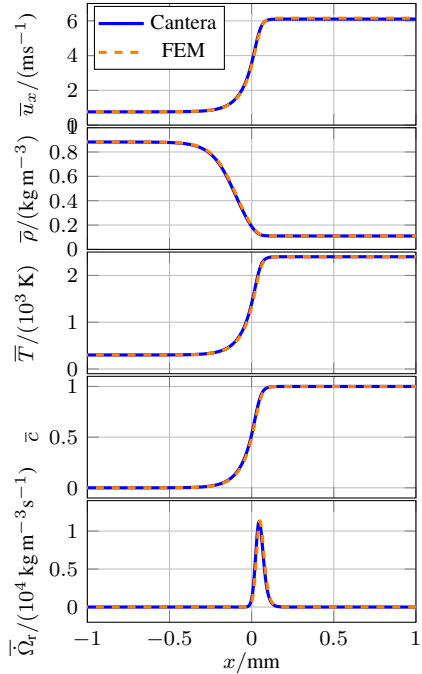

\begin{figure}[t]
    \centering
\input{figures/ZirwesSpektrum}
\caption{\gls{DR} for the 2D flame setup given by Eqs.~(\ref{eq:NonLinear:Conti}--\ref{eq:nonLinear:perfectGasLaw}).}
    \label{fig:ZirwesSpektrum}
\end{figure}
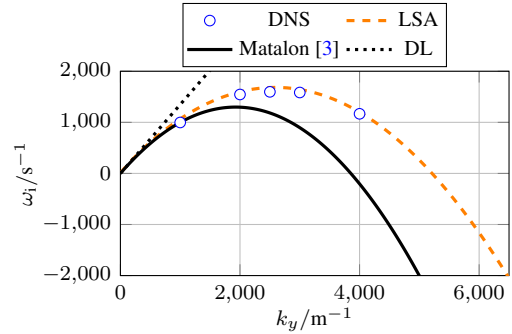

\begin{figure*}[t]
    \centering\begin{subfigure}{0.364\linewidth}
        \centering
        \input{figures/k1000}
        \caption{$k_y = 1000 \, \text{m}^{-1}$}
        \label{fig:sub1}
    \end{subfigure}
    \centering\begin{subfigure}{0.298\linewidth}
        \centering
        \input{figures/k2500}
        \caption{$k_y = 2500 \, \text{m}^{-1}$}
        \label{fig:sub1}
    \end{subfigure}
        \centering\begin{subfigure}{0.298\linewidth}
        \centering
        \input{figures/k4000}
        \caption{$k_y = 4000 \, \text{m}^{-1}$}
        \label{fig:sub1}
    \end{subfigure}
\caption{Perturbation quantities for the 2D model flame of finite thickness in velocities, pressure, temperature, progress variable and reaction rate for different wave numbers $k_y$.}
    \label{fig:ZirwesModes}
\end{figure*}
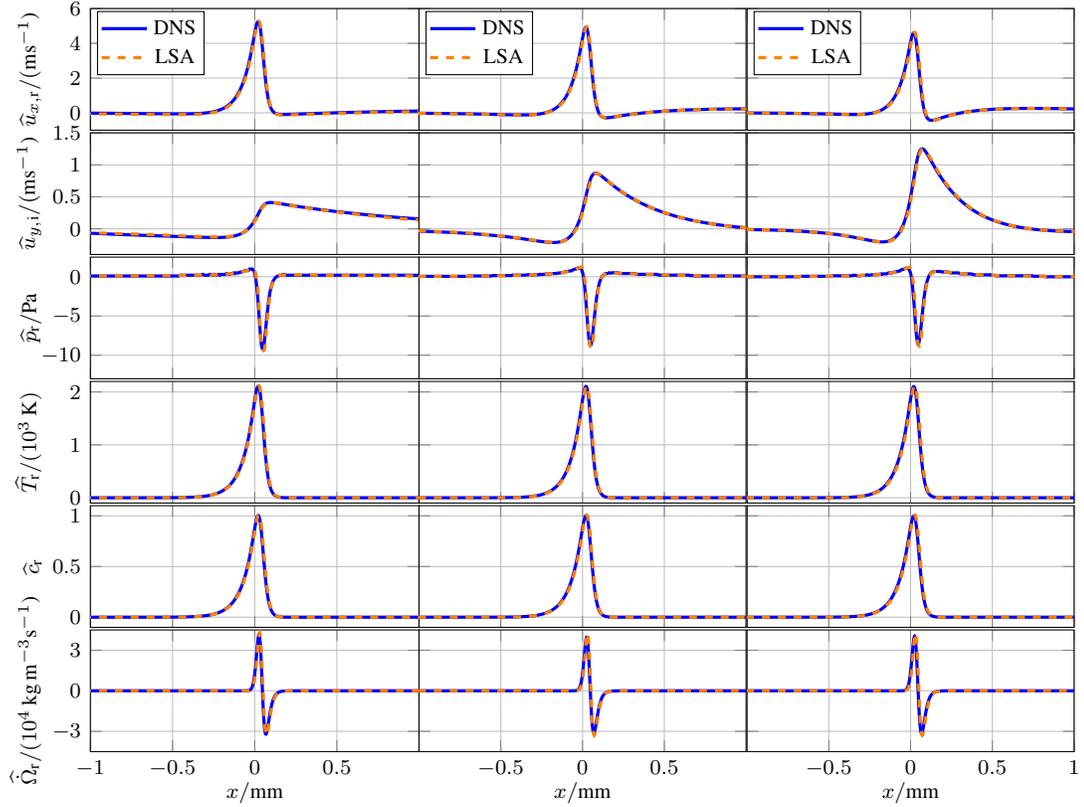
\begin{figure}[hbt]
    \centering
\input{figures/ZirwesSpektrum3D}
\caption{\gls{DR} for the 3D flame setup given by Eqs.~(\ref{eq:NonLinear:Conti})--(\ref{eq:nonLinear:perfectGasLaw}).}
    \label{fig:ZirwesSpektrum3D}
\end{figure}
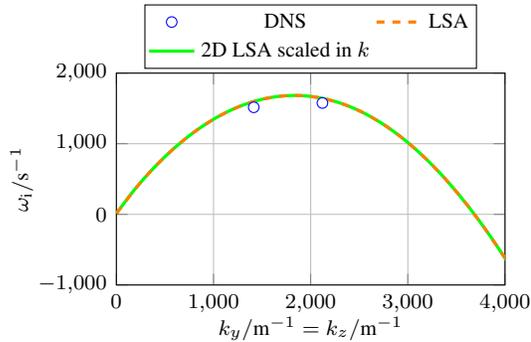
To demonstrate the applicability of the proposed framework to a more realistic flame configuration than the Darrieus-Landau configuration, the linear stability of the model flame defined by Eqs. (\ref{eq:NonLinear:Conti}--\ref{eq:nonLinear:perfectGasLaw}) is analyzed. Here, at first, a 2D flame is considered, while in the following section, the analysis will be extended to three spatial dimensions. 
In the following, the results of the \gls{GEVP}-based \gls{LSA} are compared against the corresponding \gls{DNS} results and algebraic models. 

Note that linearization around a base state obtained from either Cantera~\cite{cantera} or OpenFOAM lead to non-negligible errors, particularly in the \glspl{DR}. Only obtaining the base state on the same numerical grid and with the same numerical approach as the subsequent \gls{LSA}, i.e. Taylor-Hood \gls{CG}--\gls{FEM}, lead to the results presented in this study. This discrepancy is likely due to interpolation errors from Cantera or OpenFOAM in combination with the stiff source term in the reaction zone. The resulting \gls{FEM} base state is compared against the corresponding Cantera profiles in Fig.~\ref{fig:baseState}. 

Figure~\ref{fig:ZirwesSpektrum} presents the \gls{DR} based on different modeling techniques. The algebraic models obtained from Darrieus~\cite{darrieus1938propagation} and Landau~\cite{Landau1944} (\ref{line:DRDL}) as well as the results of Matalon et al.~\cite{Matalon1982} (\ref{line:DRMatalon}) are compared against the \gls{DNS} results, which are shown by blue circles (\ref{line:DNS}). The algebraic models remain accurate at very low wave numbers; however, deviations from the \gls{DNS} increase significantly with the wavenumber, as already discussed in many previous studies (e.g.~\cite{Berger2022a,pitsch2024transition}). Finally, the results of the \gls{GEVP}-based \gls{LSA} introduced in this study are shown by the orange dashed line (\ref{line:DRReal}). In contrast to the algebraic models, the \gls{GEVP}-based \gls{LSA} aligns very well with the \gls{DNS} results. Furthermore, the \gls{GEVP}-based \gls{LSA} analysis allows to determine the \gls{DR} at wave numbers, where the growth rate is negative at high accuracy. In contrast, determining \glspl{DR} with \gls{DNS} remains challenging and prone to significant errors when the growth rate is negative. 

The spatial distribution of the dominant eigenmodes is displayed in Fig.~\ref{fig:ZirwesModes} for three wavenumbers $k_y$. As in the results of the infinitely thin flame front, the fluctuations of all state variables is in phase, except $\widehat{u}_y$, which is shifted in phase by exactly ${\pi}/{2}$ in comparison to the remaining quantities. Therefore, Fig.~\ref{fig:ZirwesModes} illustrates the imaginary part of $\widehat{u}_y$ and the real part for all remaining fluctuations.

Figure~\ref{fig:ZirwesModes} demonstrates, that all perturbation quantities are in excellent agreement with the \gls{DNS} results. It is noteworthy, that the fluctuation in progress variable $\widehat{c}$ is largely equal for all investigated wavenumbers. This also translates to the fluctuations in reaction rate and temperature, since they are a function of the progress variable alone in the current model. The maximum fluctuation in pressure and streamwise velocity, on the other hand, seems to decrease with increasing wavenumber. The strongest dependency on the wavenumber is shown by the fluctuation in cross-streamwise velocity, which significantly increases in maximum amplitude as the wavenumber increases. 

The results demonstrate that the \gls{GEVP}-based \gls{LSA} captures both the temporal amplification and spatial structure of intrinsic laminar flame instabilities for finite-thickness reaction zones with an accuracy equal to the one of \gls{DNS}. The quantitative agreement with theoretical expectations and the robustness of the computed eigenmodes confirm that the framework provides a solid foundation for subsequent extensions toward chemically more complex and physically more realistic configurations with detailed chemistry and diffusion models. Furthermore, the 1D \gls{GEVP}-based \gls{LSA} on a single CPU core takes approximately $0.47\, \text{s}$, corresponding to a speed-up of six orders in magnitude in comparison to the \gls{DNS}. Details on the numerical costs are given in Table~\ref{tab:numCost}.

\begin{table*}[hbt]
\centering
\caption{Computational time statistics per wavenumber.}
\begin{tabular}{lcccc}
\hline
Method & \makecell{Computing time\\in seconds} & \makecell{Number of\\CPU cores} & CPU core hours & \makecell{Speed up\\against \gls{DNS}} \\
\hline
\gls{DNS} 2D & $1.4\cdot10^5$ & $8$ & $3.1\cdot10^2$ & -- \\
\gls{DNS} 3D & $5.2\cdot10^5$ & $228$ & $3.3\cdot10^4$ & -- \\
\gls{GEVP}-\gls{LSA} 2D & $0.47$ & $1$ & $1.31\cdot 10^{-4} $ & $10^6$ \\
\gls{GEVP}-\gls{LSA} 3D& $0.6$  & $1$ & $1.67\cdot 10^{-4} $ & $10^8$ \\
\hline
\end{tabular}
\label{tab:numCost}
\end{table*}

\subsection{Stability of the three-dimensional model flame with finite thickness}\label{sec:modelFlame3D}
Finally, the stability of the model flame with finite thickness in 3D space is considered. The \gls{DR} of the \gls{GEVP}-based \gls{LSA} is illustrated against \gls{DNS} results in Fig.~\ref{fig:ZirwesSpektrum3D}. As for the 2D analysis, the \gls{LSA} is in very good agreement with the \gls{DNS} results. Note that while the numerical cost for the \gls{DNS} increased by a factor of $\mathcal{O}(100)$ in comparison to 2D \gls{DNS}, the increase in the \gls{GEVP}-\gls{LSA} cost from $0.47\,\text{s}$ in two dimensions to $0.6\,\text{s}$ in three dimensions is marginal.

\begin{figure}
    \centering
    \input{figures/DRContourplot} 
    \caption{\gls{DR} of the 3D flame with finite thickness: Growth rate $\omega_i$ as a function of the wave numbers $k_y$ and $k_z$.}
    \label{fig:DR3D}
\end{figure}
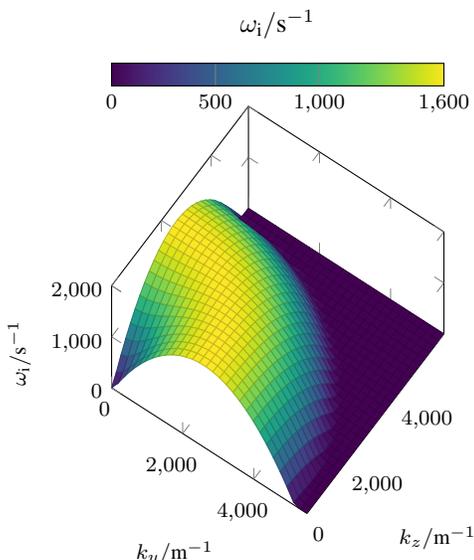

Figure~\ref{fig:DR3D} furthermore illustrates the \gls{DR} of the 2D flame from Section~\ref{sec:modelFlame2D}, where the wavenumber $k_{2D}$ is divided by $\sqrt{2}$. The results show that the scaled 2D \gls{DR} collapses onto the 3D \gls{DR}. Note, that this behavior is expected and a well known phenomenon (see e.g. Matalon~\cite{matalon_2018}). Due to the rotational symmetry in the 
$y$-$z$-plane follows that 
\begin{equation}
    k_{2D}^2 =  k_{y,3D}^2 +k_{z,3D}^2 \label{eq:2D23D}
\end{equation}
To illustrate this equation, Fig.~\ref{fig:ZirwesSpektrum3D} shows the \gls{DR} of the 3D flame as a function of $k_y$ and $k_z$ obtained by the 3D \gls{GEVP}-based \gls{LSA}, exhibiting the expected rotational symmetry around $k_y=k_z=0$. Using Eq.~(\ref{eq:2D23D}), a mapping from 2D to 3D flames can be obtained, making the explicit computation of 3D \glspl{DR} unnecessary, as long as the rotational symmetry of the problem remains intact. While this symmetry holds for planar flames with homogeneous base states, it may be broken for more complex configurations such as expanding flames, sheared flames, or flames with curvature varying along the flame front. Note that particularly in the context of \gls{LES} sub-grid scale modeling such flames are much more common in turbulent flows than planar flames. The present numerical framework opens up exploring the stability properties of such inherently 3D flame configurations. For example, inhomogeneously curved flames can be taken into consideration by discretizing the linearized governing equations Eq.~(\ref{eq:lin:conti})--(\ref{eq:lin_T}) in cylindrical and spheroid coordinate systems at a comparable low numerical cost to the \gls{GEVP}-\gls{LSA} in the present study.

\section{Conclusion\label{sec:conclusion}}\addvspace{10pt}
This study develops a \gls{GEVP}-based \gls{LSA} framework for intrinsic instabilities of laminar premixed flames. By formulating the stability problem directly in terms of the linearized reactive Navier–Stokes equations of a 1D base flame, the method enables the direct computation of \glspl{DR} and spatial eigenmodes without time integration of the nonlinear governing equations. The framework was first validated using the classical Darrieus–Landau configuration, where the numerical results accurately reproduce the analytical dispersion relation and modal structure. It was subsequently applied to a model flame of finite thickness governed by the reactive Navier–Stokes equations, where the predicted growth rates and perturbation fields show excellent agreement with corresponding direct numerical simulations.

The results demonstrate that the proposed GEVP-LSA approach provides DNS-level accuracy while reducing computational cost by up to eight orders of magnitude. This efficiency enables systematic stability investigations across wide parameter ranges that are difficult to access using DNS-based dispersion-relation analysis. Because the method operates directly on the governing equations rather than simplified flame front models, it provides a general and extensible framework for analyzing intrinsic flame instabilities in reacting flows. In particular, the formulation provides a foundation for incorporating additional physical effects such as thermodiffusion, detailed transport models, and multi-species chemistry, and for investigating more complex flame configurations relevant to combustion modeling. The present results therefore establish GEVP-LSA as an efficient and scalable tool for studying intrinsic flame instabilities and for supporting the development of physics-based subgrid models in large-eddy simulations of reacting flows in the future.

\acknowledgement{CRediT authorship contribution statement} \addvspace{10pt}

{\bf Thomas Ludwig Kaiser}: Conceptualization, Methodology, Software (Non-linear \gls{FEM} calculation and \gls{LSA}), Formal analysis, Writing - Original Draft, Funding acquisition. {\bf Peter Munch}: Methodology, Writing - Review and Editing. {\bf Sandra May}: Methodology, Writing - Review and Editing. {\bf Thorsten Zirwes}: Software, Data Curation, Writing - Review and Editing.

\acknowledgement{Declaration of competing interest} \addvspace{10pt}

The authors declare that they have no known competing financial interests or personal relationships that could have appeared to influence the work reported in this paper.

\acknowledgement{Acknowledgments} \addvspace{10pt}

Funded by the Deutsche Forschungsgemeinschaft (DFG, German Research Foundation) – 530442286, 523881008. The authors gratefully acknowledge the computing time provided on the high-performance computer HoreKa by the National High-Performance Computing Center at KIT~(NHR@KIT). This center is jointly supported by the Federal Ministry of Education and Research and the Ministry of Science, Research and the Arts of Baden-Württemberg, as part of the National High-Performance Computing~(NHR) joint funding program (\url{https://www.nhr-verein.de/en/our-partners}). HoreKa is partly funded by the German Research Foundation~(DFG).

\footnotesize
\baselineskip 9pt

\clearpage
\thispagestyle{empty}
\bibliographystyle{proci}
\bibliography{PROCI_LaTeX}


\newpage

\small
\baselineskip 10pt


\end{document}

%% file: acronyms.tex
\usepackage{glossaries-extra}
\setabbreviationstyle[acronym]{long-short}
\glssetcategoryattribute{acronym}{nohyperfirst}{true}

\newacronym{ATF}{ATF}{Artificially Thickened Flame}
\newacronym{CFD}{CFD}{Computational Fluid Dynamics}
\newacronym{CFL}{CFL}{Courant-Friedrichs-Lewy}
\newacronym{CG}{CG}{Continuous Galerkin}
\newacronym{DG}{DG}{Discontinuous Galerkin}
\newacronym{DNS}{DNS}{Direct Numerical Simulation}
\newacronym[\glslongpluralkey={dispersion relation}]{DR}{DR}{dispersion relation}

\newacronym{EBU}{EBU}{Eddy Break Up}
\newacronym{EP}{EWP}{Eigenwert Problem}
\newacronym{FDF}{FDF}{Flame Describing Function}
\newacronym{FE}{FE}{Finite Elemente}
\newacronym{FEM}{FEM}{Finite Element Method}
\newacronym{FES}{EXP}{Chair of Fluid Dynamics}
\newacronym{FELiCS}{FELiCS}{Finite Element Linearized Combustion Solver}

\newacronym{FSD}{FSD}{Flame Surface Density}
\newacronym{FTF}{FTF}{Flame Transfer Function}
\newacronym{GEVP}{GEVP}{Generalized Eigenvalue Problem}
\newacronym{HPC}{HPC}{High Performance Computing}
\newacronym{IOA}{IOA}{Input-Output Analysis}
\newacronym{IDP}{IDP}{Inverse Design Procedure}
\newacronym{ISTA}{ISTA}{Institut of Fluid Dynamics and Technical Acoustics}
\newacronym{IWB}{IWB}{Institute for Machine Tools and Industrial Management}
\newacronym{KH}{KH}{Kelvin-Helmholtz}
\newacronym{LA}{LA}{Linear Analysis}
\newacronym{LW}{LW}{Lax-Wendroff}
\newacronym{LES}{LES}{Large Eddy Simulation}
\newacronym{LHS}{LHS}{left hand side}
\newacronym{LNSE}{LNSE}{Lineaerized Navier-Stokes Equations}
\newacronym{LMFA}{LMFA}{Linearized Mean Field Analysis}
\newacronym{LSA}{LSA}{Linear Stability Analysis}
\newacronym{MMC}{MMC}{Multiple Mapping Conditioning}
\newacronym{NOx}{NOx}{Nitrogen oxides}
\newacronym{NSCBC}{NSCBC}{Navier-Stokes Characteristic Boundary Conditions}
\newacronym{PDE}{PDE}{Partial Differential Equation}
\newacronym{PBF-LB/M}{PBF-LB/M}{powder bed fusion of metals using a laser beam}
\newacronym{PLIF}{PLIF}{planar laser-induced fluorescence}
\newacronym{PI}{PI}{Principle Investigator}
\newacronym{PIV}{PIV}{Particle Image Velocimetry}
\newacronym{PINN}{PINN}{Physics Informed Neural Network}
\newacronym{PVC}{PVC}{Precessing Vortex Core}
\newacronym{RA}{RA}{Resolvent Analysis}
\newacronym{RANS}{RANS}{Reynolds-averaged Navier Stokes}
\newacronym{RHS}{RHS}{right hand side}
\newacronym{RST}{RST}{Reynolds-Stress Transport}
\newacronym{RQ}{RQ}{Research Question}
\newacronym{SGS}{SGS}{Subgrid Scale}
\newacronym{SPL}{SPL}{Sound Pressure Level}
\newacronym{SPP}{SPP}{Priority Program}
\newacronym{SPOD}{SPOD}{Spectral Proper Orthogonal Decomposition}
\newacronym{eSPOD}{eSPOD}{extended Spectral Proper Orthogonal Decomposition}

\newacronym{TAI}{TAI}{Thermoacoustic Instabilities}
\newacronym{TD}{TD}{thermodiffusive}
\newacronym[\glslongpluralkey={Thermodiffusive Instabilitäten}]{TDI}{TDI}{Thermodiffusive Instabilität}
\newacronym{TF}{TF}{Transfer Function}
\newacronym{TFC}{TFC}{Turbulent Flame-speed Closure}
\newacronym{TKE}{TKE}{Turbulent-Kinetic-Energy}
\newacronym{TTGC}{TTGC}{Two-Step Taylor Galerkin Scheme C}
\newacronym{TUB}{TUB}{TU Berlin}
\newacronym{TUM}{TUM}{TU Munich}
\newacronym{UCI}{UCI}{University of California Irvine}
\newacronym{WA}{WA}{work area}
\newacronym{Wasserstoff}{$\text{H}_2$}{Wasserstoff}

\newacronym{WHI}{WHI}{Wiener-Hopf Inversion}
\newacronym{felics}{FELiCS}{Finite Element Linear Combustion Solver}
\newacronym{fenics}{FEniCS}{Finite Element Computational Software}
\newacronym{TFM}{TFM}{Thickened Flame Model}

%% file: figures/LandaMean.tex
\pgfplotsset{compat=1.18,
        label style={font=\footnotesize},
        tick label style={font=\footnotesize}}
\begin{tikzpicture}
\begin{axis}[
    group style={
        group name=my plots,
        group size=5 by 1,
        horizontal sep=30pt,  
    },
    width=0.5\textwidth,
    height=4cm,
    grid=both,
    xlabel = $x$,
    xmin=-2,xmax=2,
    ymin=-1.7,ymax=1.1,
    legend style={font=\footnotesize},
    y label style={yshift=-5pt},
    grid=both,
        ylabel style={align=center},
        ylabel={$\overline{u}^*_x, \, \overline{\rho}^*$},
        xlabel=$x^*$,
        xmin = -1,
        xmax=1,
        ymin = 0,
        ymax=2.2,
        xtick = {-1,0,1},
        xticklabels = {$-\infty$,0,$+\infty$},
        legend pos=outer north east,
        legend cell align={left},
        legend style={
            at={(0.01,0.95)},
            anchor=north west,
            legend columns=1,
            font=\footnotesize,}
]
        \addplot[thick, black] coordinates 
        {
            (-1,1)
            (0,1)
            (0,2)  
            (1,2)  
        };
        \addlegendentry{$\overline{u}^*_x$}
        \label{linestyle:uxMean}
                \addplot[very thick,dashed, green] coordinates 
        {
            (-1,1)
            (0,1)
            (0,0.5)  
            (1,0.5)  
        };
        \addlegendentry{$\overline{\rho}^*$}
        \label{linestyle:rhoMean}
\end{axis}
\end{tikzpicture}

%% file: figures/spectrumLandau.tex
\pgfplotsset{compat=1.18,
        label style={font=\footnotesize},
        tick label style={font=\footnotesize}}
\begin{tikzpicture}
\begin{groupplot}[
    group style={
        group name=my plots,
        group size=2 by 1,
        horizontal sep=30pt,  
    },
    width=0.25\textwidth,
    height=4cm,
    grid=both,
    xlabel = $x$,
    xmin=-2,xmax=2,
    ymin=-1.7,ymax=1.1,
    legend style={font=\small},
    y label style={yshift=-5pt},
]

\nextgroupplot[
        grid=both,
        ylabel style={align=center},
        ylabel={$\omega^*_\text{i}$},
        xlabel=$\omega^*_\text{r}$,
                xmin=-15, xmax=15.0,
        width = 0.34\textwidth,
        ymin=-4, ymax=3.99,        
]
        \addplot[dashed, forget plot] coordinates 
        {
            (0,3.099407093782345)
            (15,3.099407093782345)    
        };

        \addplot [
            color=blue,
            fill=none,
            mark=*,
            mark options={fill=white,scale=1.5,},
            only marks
        ] table [x=omega_r, y=omega_i, col sep=comma] {figures/Data/spectrum_analytical.csv}; \label{line:analyticSpectrum}

        \addplot [
            very thick,
            color=orange,
            mark=star,
            mark options={fill=orange,scale=1},
            only marks
        ] table [x=omega_r, y=omega_i, col sep=comma] {figures/Data/spectrum_unstable.csv};
        \label{line:numSpectrumPhysical}

        \addplot [
            color=black,
            mark=x,
            mark options={fill=orange,scale=1},
            only marks
        ] table [x=omega_r, y=omega_i, col sep=comma] {figures/Data/spectrum.csv};
        \label{line:numSpectrumSpurious}
\nextgroupplot[
        grid=both,
        ylabel style={align=center},
        ylabel={},
        xlabel=$k^*$,
        xmin = 0,
        xmax=10,
        ymin = -4,
        ymax=3.99,
        width = 0.25\textwidth,
        yticklabels={,},
        at={(my plots c1r1.east)}, anchor=west,
        legend pos=outer north east,
        legend cell align={left},
        legend style={
            at={(0.95,0.05)},
            anchor=south east,
            legend columns=1,
            font=\footnotesize,
        }
]
       \addplot [
            color=blue, 
            very thick,
            forget plot,
        ] table [x=k, y=omega_analyt, col sep=comma] {figures/Data/dispersion_relation.csv}; \label{line:analytical}

        \addplot [
            color=orange,
            very thick,
            dashed,
            forget plot,
        ] table [x=k, y=omega_DG, col sep=comma] {figures/Data/dispersion_relation.csv}; \label{line:numerical}
        \addplot[dashed, forget plot] coordinates 
        {
            (8,-4)
            (8,3.099407093782345)
            (0,3.099407093782345)    
        };
    \addlegendimage{color=blue, 
            thin,  
            fill=none,
            mark=*,
            mark options={fill=white,scale=2,},} \label{line:LandauAll}

    \addlegendimage{orange, dashed, very thick,
            mark=star,
            mark options={fill=orange,scale=1.5},}\label{line:numAll}
    \addlegendimage{ color=black,
            mark=x,
            mark options={fill=orange,scale=2},
            only marks} \label{line:Spurious}
\end{groupplot}
\end{tikzpicture}

%% file: figures/LandauModes.tex
\pgfplotsset{compat=1.18,
        label style={font=\footnotesize},
        tick label style={font=\footnotesize},xlabel style={yshift=3pt}}
\begin{tikzpicture}
\begin{groupplot}[
    group style={
        group name=my plots,
        group size=1 by 3 ,
        horizontal sep=30pt,  
        vertical sep=1pt,
    },
    width=0.25\textwidth,
    grid=both,
    xmin=-4, xmax=4,
    height=0.55\linewidth,
    width =1\linewidth,
    xmin=-2,xmax=2,
    ymin=-1.7,ymax=1.1,
    legend style={font=\small},
    y label style={yshift=-5pt},
]
\nextgroupplot[
    ylabel = {$\widehat{u}^*_x$}, 
        ymin=-0.4,ymax=1.1,
            xticklabels={,},
    xtick={},
    legend style={
    font=\footnotesize,
        at={(0.02,0.98)}, 
        anchor=north west,  
inner sep=1pt,
        legend columns=1 
    },
]
\addplot [
            color= blue,
             very thick,,
        ] table [x=x, y=ux_real, col sep=comma] {figures/Data/velocities_analytical.csv};
\addplot [
            color=orange,
            very thick,
            dashed
        ] table [x=x, y=ux_real, col sep=comma] {figures/Data/velocities.csv};
        \legend{analytic, numerical}
\addplot [
            color=blue,
             very thick,,
        ] table [x=x, y=ux_imag, col sep=comma] {figures/Data/velocities_analytical.csv};
\addplot [
            color=orange,
            very thick,
            dashed
        ] table [x=x, y=ux_imag, col sep=comma] {figures/Data/velocities.csv};
\node[name=real, anchor=south,font=\footnotesize] at (axis cs:1,0.7) {real};
\draw (real.west) -- (axis cs:0,1);
\node[name=imag, anchor=north,font=\footnotesize] at (axis cs:1,0.7) {imag};
\draw (imag.south) -- (axis cs:0.5,0);

\nextgroupplot[
    ylabel = {$\widehat{u}^*_y$}, 
        ymin=-1.7,ymax=1.1,
            xticklabels={,},
    xtick={},
]
\addplot [
            color= blue,
             very thick,,
        ] table [x=x, y=uy_real, col sep=comma] {figures/Data/velocities_analytical.csv};
\addplot [
            color=orange,
            very thick,
            dashed
        ] table [x=x, y=uy_real, col sep=comma] {figures/Data/velocities.csv};
\addplot [
            color=blue,
            very thick,
        ] table [x=x, y=uy_imag, col sep=comma] {figures/Data/velocities_analytical_minus.csv};
\addplot [
            color=blue,
             very thick,
        ] table [x=x, y=uy_imag, col sep=comma] {figures/Data/velocities_analytical_plus.csv};
\addplot [
            color=orange,
            very thick,
            dashed
        ] table [x=x, y=uy_imag, col sep=comma] {figures/Data/velocities_minus.csv};
        \addplot [
            color=orange,
            very thick,
            dashed
        ] table [x=x, y=uy_imag, col sep=comma] {figures/Data/velocities_plus.csv};

\node[name=real, anchor=south,font=\footnotesize] at (axis cs:-1,-0.6) {real};
\draw (real.east) -- (axis cs:-0.1,0);
\node[name=imag1, anchor=north,font=\footnotesize] at (axis cs:1,1) {imag};
\draw (imag1.west) -- (axis cs:0,1);
\node[name=imag2, anchor=north,font=\footnotesize] at (axis cs:-1,-0.6) {imag};
\draw (imag2.east) -- (axis cs:0,-1.4);

\nextgroupplot[
    ylabel = {$\widehat{p}^*$}, 
        ymin=-1.5,ymax=0.1,
    xlabel = {$x^*$},
]
\addplot [
            color= blue,
             very thick,,
        ] table [x=x, y=p_real, col sep=comma] {figures/Data/pressure_analytical.csv};
\addplot [
            color=orange,
            very thick,
            dashed
        ] table [x=x, y=p_real, col sep=comma] {figures/Data/pressure.csv};
\addplot [
            color=blue,
             very thick,,
        ] table [x=x, y=p_imag, col sep=comma] {figures/Data/pressure_analytical.csv};
\addplot [
            color=orange,
            very thick,
            dashed
        ] table [x=x, y=ux_imag, col sep=comma] {figures/Data/velocities.csv};
\node[name=real, anchor=south,font=\footnotesize] at (axis cs:1,-1) {real};
\draw (real.west) -- (axis cs:0.25,0);
\node[name=imag, anchor=north,font=\footnotesize] at (axis cs:-1,-1) {imag};
\draw (imag.east) -- (axis cs:-0.06,-0.9);
\end{groupplot}
\end{tikzpicture}

%% file: figures/baseState.tex
\pgfplotsset{compat=1.18,
        label style={font=\footnotesize},
        tick label style={font=\footnotesize}}
\begin{tikzpicture}
\begin{groupplot}[
    group style={
        group name=my plots,
        group size=1 by 6 ,
        horizontal sep=30pt,  
        vertical sep=1pt,
    },
    width=5.95cm,
    height=3.2cm,
    grid=both,
    xmin=-1, xmax=1,
    ymin=-1.7,ymax=1.1,
    legend style={font=\small},
    ylabel style={
    at={(axis description cs:0,0.5)},
    anchor=south,
    yshift=15pt
},
]
\nextgroupplot[
    ylabel = {$\overline{u}_{x} / (\text{ms}^{-1}$}), 
        ymin=0,ymax=7,
            xticklabels={,},
    xtick={},    legend style={
        font=\footnotesize,
        at={(0.02,0.98)}, 
        anchor=north west,  
inner sep=1pt,
        legend columns=1 
    },
]
\addplot [
    color=blue,
     very thick,
    solid,smooth
]
table [
    x expr=\thisrow{x}*1000,    
    y expr=\thisrow{ux}, 
    col sep=comma
] {figures/Data/baseState_cantera.csv};

\addplot [
    color=orange,
    very thick,
    dashed,smooth
]
table [
    x expr=\thisrowno{0}*1000,    
    y expr=\thisrowno{1}, 
    col sep=comma
] {figures/Data/baseState_FEM.csv};
        \legend{Cantera, \gls{FEM}}
\nextgroupplot[
    ylabel = {$\overline{\rho}/ (\text{kg}\,\text{m}^{-3})$}, 
        ymin=0,ymax=0.999,
            xticklabels={,},
    xtick={},
]
\addplot [
    color=blue,
     very thick,
    solid,smooth
]
table [
    x expr=\thisrow{x}*1000,    
    y expr=\thisrow{rho}, 
    col sep=comma
] {figures/Data/baseState_cantera.csv};

\addplot [
    color=orange,
    very thick,
    dashed,smooth
]
table [
    x expr=\thisrowno{0}*1000,    
    y expr=\thisrow{rho}, 
    col sep=comma
] {figures/Data/baseState_FEM.csv};
\nextgroupplot[
    ylabel = {$\overline{T}/(10^3 \, \text{K})$}, 
        ymin=-0.1,ymax=2.5,
            xticklabels={,},
    xtick={},
]
\addplot [
    color=blue,
     very thick,
    solid,smooth
]
table [
    x expr=\thisrow{x}*1000,    
    y expr=\thisrow{T}/1000, 
    col sep=comma
] {figures/Data/baseState_cantera.csv};

\addplot [
    color=orange,
    very thick,
    dashed,smooth
]
table [
    x expr=\thisrowno{0}*1000,    
    y expr=\thisrow{T}/1000, 
    col sep=comma
] {figures/Data/baseState_FEM.csv};
\nextgroupplot[
    ylabel = {$\overline{c}$}, 
        ymin=-0.1,ymax=1.1,
            xticklabels={,},
    xtick={},
]
\addplot [
    color=blue,
     very thick,
    solid,smooth
]
table [
    x expr=\thisrow{x}*1000,    
    y expr=\thisrow{c}, 
    col sep=comma
] {figures/Data/baseState_cantera.csv};

\addplot [
    color=orange,
    very thick,
    dashed,smooth
]
table [
    x expr=\thisrowno{0}*1000,    
    y expr=\thisrow{c}, 
    col sep=comma
] {figures/Data/baseState_FEM.csv};
\nextgroupplot[
    ylabel = {$\overline{\dot{\Omega}}_\text{r}/ (10^{4}\,\text{kg}\,\text{m}^{-3}\text{s}^{-1}$}), 
        ymin=-0.2,ymax=1.4,
            xlabel={$x/\text{mm}$},xlabel style={yshift=5pt}
]
\addplot [
    color=blue,
     very thick,
    solid,smooth
]
table [
    x expr=\thisrow{x}*1000,    
    y expr=\thisrow{RR}/10000, 
    col sep=comma
] {figures/Data/baseState_cantera.csv};

\addplot [
    color=orange,
    very thick,
    dashed,smooth,
]
table [
    x expr=\thisrowno{0}*1000,    
    y expr=\thisrow{RR}/10000, 
    col sep=comma
] {figures/Data/baseState_FEM.csv};
\end{groupplot}
\end{tikzpicture}

%% file: figures/ZirwesSpektrum.tex
\pgfplotsset{compat=1.18,
        label style={font=\footnotesize},
        tick label style={font=\footnotesize}}
\begin{tikzpicture}
\begin{groupplot}[
    group style={
        group name=my plots,
        group size=2 by 1,
        horizontal sep=30pt,  
    },
    width=1.0\linewidth,
    height=4.3cm,
    grid=both,
    xlabel = $x$,
    xmin=0,xmax=6500,
    ymin=-1.7,ymax=1.1,
    legend style={font=\small},
    y label style={yshift=-5pt},
]

\nextgroupplot[
        grid=both,
        xlabel style={yshift=5pt},
        ylabel style={align=center},
        ylabel={$\omega_\text{i}/\text{s}^{-1}$},
        xlabel=$k_y/\text{m}^{-1}$,
        ymin=-2000, ymax=2000,    
        legend columns = 2,
            legend style={
                font=\footnotesize,
        at={(0.5,1.02)}, 
        anchor=south,  
inner sep=1pt,
        legend columns=1 
    },
]

        \addplot [
            color=blue,
            fill=none,
            mark=*,
            mark options={fill=white,scale=1,},
            only marks
        ] coordinates 
        {

            (1000,997)
            (2000,1545)
            (2500,1597)
            (3000,1584)
            (4000,1167)
            
        };
        \addlegendentry{\gls{DNS}};\label{line:DNS}

        \addplot [
            very thick,
            color=orange,
            very thick, dashed,smooth,
        ] table [x=k, y=omega_i, col sep=comma] {figures/Data/ArrheniusDR_2D_9000.csv};
        \label{line:DRReal}
        \addlegendentry{\gls{LSA}};

                \addplot [
            very thick,
            color=black,
            domain=0:6000,
            samples=100,
           very thick, smooth
        ] {-0.000349660070138458*x*x + 1.34754871560497*x};
        \label{line:DRMatalon}
        \addlegendentry{Matalon~\cite{Matalon2003}};

        \addplot [
            very thick,
            color=black,
            domain=0:4000,
            samples=10,
           very thick, dotted, smooth
        ] {1.34754871560497*x};
        \label{line:DRDL}
        \addlegendentry{DL};

\end{groupplot}
\end{tikzpicture}

%% file: figures/k1000.tex
\pgfplotsset{compat=1.18,
        label style={font=\footnotesize},
        tick label style={font=\footnotesize}}
\begin{tikzpicture}
\begin{groupplot}[
    group style={
        group name=my plots,
        group size=1 by 6 ,
        horizontal sep=30pt,  
        vertical sep=1pt,
    },
    width=5.95cm,
    height=3.2cm,
    grid=both,
    xmin=-1, xmax=1,
    ymin=-1.7,ymax=1.1,
    legend style={font=\small},
    ylabel style={
    at={(axis description cs:0,0.5)},
    anchor=south,
    yshift=15pt
},
    xtick={-1, -0.5,0,0.5}
]
\nextgroupplot[
    ylabel = {$\widehat{u}_{x,\text{r}} / (\text{ms}^{-1}$}), 
        ymin=-1,ymax=6,
            xticklabels={,},
    xtick={},    legend style={
        font=\footnotesize,
        at={(0.02,0.98)}, 
        anchor=north west,  
inner sep=1pt,
        legend columns=1 
    },
]
\addplot [
    color=blue,
     very thick,
    solid,smooth
]
table [
    x expr=\thisrow{x}*1000,    
    y expr=\thisrow{ux}, 
    col sep=comma
] {figures/Data/flucs_k1001.csv};

\addplot [
    color=orange,
    very thick,
    dashed,smooth
]
table [
    x expr=\thisrowno{0}*1000,    
    y expr=\thisrowno{1}, 
    col sep=comma
] {figures/Data/eigenvectors/beta1001_Ux.csv};
        \legend{\gls{DNS}, \gls{LSA}}
\nextgroupplot[
    ylabel = {$\widehat{u}_{y,\text{i}}/ (\text{ms}^{-1}$}), 
        ymin=-0.4,ymax=1.5,
            xticklabels={,},
    xtick={},
]
\addplot [
    color=blue,
     very thick,
    solid,smooth
]
table [
    x expr=\thisrow{x}*1000,    
    y expr=\thisrow{uy}*-1, 
    col sep=comma
] {figures/Data/flucs_k1001.csv};

\addplot [
    color=orange,
    very thick,
    dashed,smooth
]
table [
    x expr=\thisrowno{0}*1000,    
    y expr=\thisrowno{2}, 
    col sep=comma
] {figures/Data/eigenvectors/beta1001_Uy.csv};
\nextgroupplot[
    ylabel = {$\widehat{p}_\text{r}/ \text{Pa}$}, 
        ymin=-13,ymax=2.5,
            xticklabels={,},
    xtick={},
]
\addplot [
    color=blue,
     very thick,
    solid,smooth
]
table [
    x expr=\thisrow{x}*1000,    
    y expr=\thisrow{p}, 
    col sep=comma
] {figures/Data/flucs_k1001.csv};

\addplot [
    color=orange,
    very thick,
    dashed,smooth
]
table [
    x expr=\thisrowno{0}*1000,    
    y expr=\thisrowno{1}, 
    col sep=comma
] {figures/Data/eigenvectors/beta1001_p.csv};
\nextgroupplot[
    ylabel = {$\widehat{T}_\text{r}/(10^3 \, \text{K})$}, 
        ymin=-0.1,ymax=2.2,
            xticklabels={,},
    xtick={},
]
\addplot [
    color=blue,
     very thick,
    solid,smooth
]
table [
    x expr=\thisrow{x}*1000,    
    y expr=\thisrow{T}/1000, 
    col sep=comma
] {figures/Data/flucs_k1001.csv};

\addplot [
    color=orange,
    very thick,
    dashed,smooth
]
table [
    x expr=\thisrowno{0}*1000,    
    y expr=\thisrowno{1}/1000, 
    col sep=comma
] {figures/Data/eigenvectors/beta1001_T.csv};
\nextgroupplot[
    ylabel = {$\widehat{c}_\text{r}$}, 
        ymin=-0.1,ymax=1.1,
            xticklabels={,},
    xtick={},
]
\addplot [
    color=blue,
     very thick,
    solid,smooth
]
table [
    x expr=\thisrow{x}*1000,    
    y expr=\thisrow{c}, 
    col sep=comma
] {figures/Data/flucs_k1001.csv};

\addplot [
    color=orange,
    very thick,
    dashed,smooth
]
table [
    x expr=\thisrowno{0}*1000,    
    y expr=\thisrowno{1}, 
    col sep=comma
] {figures/Data/eigenvectors/beta1001_c.csv};
\nextgroupplot[
    ylabel = {$\widehat{\dot{\Omega}}_\text{r}/(10^{4}\,\text{kg}\,\text{m}^{-3}\text{s}^{-1}$}), 
        ymin=-4.5,ymax=4.5,
        ytick={-3,0,3},
            xlabel={$x/\text{mm}$},xlabel style={yshift=5pt}
]
\addplot [
    color=blue,
     very thick,
    solid,smooth
]
table [
    x expr=\thisrow{x}*1000,    
    y expr=\thisrow{RR}/10000, 
    col sep=comma
] {figures/Data/flucs_k1001.csv};

\addplot [
    color=orange,
    very thick,
    dashed,smooth,
]
table [
    x expr=\thisrowno{0}*1000,    
    y expr=\thisrowno{1}/10000, 
    col sep=comma
] {figures/Data/eigenvectors/beta1001_source.csv};
\end{groupplot}
\end{tikzpicture}

%% file: figures/k2500.tex
\pgfplotsset{compat=1.18,
        label style={font=\footnotesize},
        tick label style={font=\footnotesize}}
\begin{tikzpicture}
\begin{groupplot}[
    group style={
        group name=my plots,
        group size=1 by 6 ,
        horizontal sep=30pt,  
        vertical sep=1pt,
    },
    width=5.93cm,
    height=3.2cm,
    grid=both,
    xmin=-1, xmax=1,
    ymin=-1.7,ymax=1.1,
    legend style={font=\small},
    y label style={yshift=-5pt},
    yticklabels={,},
    xtick={-0.5,0,0.5}
]
\nextgroupplot[
        ymin=-1,ymax=6,
            xticklabels={,},
    xtick={},    legend style={
        font=\footnotesize,
        at={(0.02,0.98)}, 
        anchor=north west,  
inner sep=1pt,
        legend columns=1 
    },
]
\addplot [
    color=blue,
     very thick,
    solid,smooth
]
table [
    x expr=\thisrow{x}*1000,    
    y expr=\thisrow{ux}, 
    col sep=comma
] {figures/Data/flucs_k2506.csv};

\addplot [
    color=orange,
    very thick,
    dashed,smooth
]
table [
    x expr=\thisrowno{0}*1000,    
    y expr=\thisrowno{1}, 
    col sep=comma
] {figures/Data/eigenvectors/beta2501_Ux.csv};
        \legend{\gls{DNS}, \gls{LSA}}
\nextgroupplot[
        ymin=-0.4,ymax=1.5,
            xticklabels={,},
    xtick={},
]
\addplot [
    color=blue,
     very thick,
    solid,smooth
]
table [
    x expr=\thisrow{x}*1000,    
    y expr=\thisrow{uy}*-1, 
    col sep=comma
] {figures/Data/flucs_k2506.csv};

\addplot [
    color=orange,
    very thick,
    dashed,smooth
]
table [
    x expr=\thisrowno{0}*1000,    
    y expr=\thisrowno{2}, 
    col sep=comma
] {figures/Data/eigenvectors/beta2501_Uy.csv};
\nextgroupplot[
        ymin=-13,ymax=2.5,
            xticklabels={,},
    xtick={},
]
\addplot [
    color=blue,
     very thick,
    solid,smooth
]
table [
    x expr=\thisrow{x}*1000,    
    y expr=\thisrow{p}, 
    col sep=comma
] {figures/Data/flucs_k2506.csv};

\addplot [
    color=orange,
    very thick,
    dashed,smooth
]
table [
    x expr=\thisrowno{0}*1000,    
    y expr=\thisrowno{1}, 
    col sep=comma
] {figures/Data/eigenvectors/beta2501_p.csv};
\nextgroupplot[
        ymin=-100,ymax=2200,
            xticklabels={,},
    xtick={},
]
\addplot [
    color=blue,
     very thick,
    solid,smooth
]
table [
    x expr=\thisrow{x}*1000,    
    y expr=\thisrow{T}, 
    col sep=comma
] {figures/Data/flucs_k2506.csv};

\addplot [
    color=orange,
    very thick,
    dashed,smooth
]
table [
    x expr=\thisrowno{0}*1000,    
    y expr=\thisrowno{1}, 
    col sep=comma
] {figures/Data/eigenvectors/beta2501_T.csv};
\nextgroupplot[
        ymin=-0.1,ymax=1.1,
            xticklabels={,},
    xtick={},
]
\addplot [
    color=blue,
     very thick,
    solid,smooth
]
table [
    x expr=\thisrow{x}*1000,    
    y expr=\thisrow{c}, 
    col sep=comma
] {figures/Data/flucs_k2506.csv};

\addplot [
    color=orange,
    very thick,
    dashed,smooth
]
table [
    x expr=\thisrowno{0}*1000,    
    y expr=\thisrowno{1}, 
    col sep=comma
] {figures/Data/eigenvectors/beta2501_c.csv};
\nextgroupplot[
        ymin=-4.5,ymax=4.5,
        ytick={-3,0,3},
            xlabel={$x/\text{mm}$},xlabel style={yshift=5pt},
]
\addplot [
    color=blue,
     very thick,
    solid,smooth
]
table [
    x expr=\thisrow{x}*1000,    
    y expr=\thisrow{RR}/10000, 
    col sep=comma
] {figures/Data/flucs_k2506.csv};

\addplot [
    color=orange,
    very thick,
    dashed,smooth,
]
table [
    x expr=\thisrowno{0}*1000,    
    y expr=\thisrowno{1}/10000, 
    col sep=comma
] {figures/Data/eigenvectors/beta2501_source.csv};
\end{groupplot}
\end{tikzpicture}

%% file: figures/k4000.tex
\pgfplotsset{compat=1.18,
        label style={font=\footnotesize},
        tick label style={font=\footnotesize}}
\begin{tikzpicture}
\begin{groupplot}[
    group style={
        group name=my plots,
        group size=1 by 6 ,
        horizontal sep=30pt,  
        vertical sep=1pt,
    },
    width=5.93cm,
    height=3.2cm,
    grid=both,
    xmin=-1, xmax=1,
    ymin=-1.7,ymax=1.1,
    legend style={font=\small},
    y label style={yshift=-5pt},
    yticklabels={,},
    xtick={-0.5,0,0.5,1}
]
\nextgroupplot[
        ymin=-1,ymax=6,
            xticklabels={,},
    xtick={},    legend style={
        font=\footnotesize,
        at={(0.02,0.98)}, 
        anchor=north west,  
inner sep=1pt,
        legend columns=1 
    },
]
\addplot [
    color=blue,
     very thick,
    solid,smooth
]
table [
    x expr=\thisrow{x}*1000,    
    y expr=\thisrow{ux}, 
    col sep=comma
] {figures/Data/flucs_k4016.csv};

\addplot [
    color=orange,
    very thick,
    dashed,smooth
]
table [
    x expr=\thisrowno{0}*1000,    
    y expr=\thisrowno{1}, 
    col sep=comma
] {figures/Data/eigenvectors/beta4000_Ux.csv};
        \legend{\gls{DNS}, \gls{LSA}}
\nextgroupplot[
        ymin=-0.4,ymax=1.5,
            xticklabels={,},
    xtick={},
]
\addplot [
    color=blue,
     very thick,
    solid,smooth
]
table [
    x expr=\thisrow{x}*1000,    
    y expr=\thisrow{uy}*-1, 
    col sep=comma
] {figures/Data/flucs_k4016.csv};

\addplot [
    color=orange,
    very thick,
    dashed,smooth
]
table [
    x expr=\thisrowno{0}*1000,    
    y expr=\thisrowno{2}, 
    col sep=comma
] {figures/Data/eigenvectors/beta4000_Uy.csv};
\nextgroupplot[
        ymin=-13,ymax=2.5,
            xticklabels={,},
    xtick={},
]
\addplot [
    color=blue,
     very thick,
    solid,smooth
]
table [
    x expr=\thisrow{x}*1000,    
    y expr=\thisrow{p}, 
    col sep=comma
] {figures/Data/flucs_k4016.csv};

\addplot [
    color=orange,
    very thick,
    dashed,smooth
]
table [
    x expr=\thisrowno{0}*1000,    
    y expr=\thisrowno{1}, 
    col sep=comma
] {figures/Data/eigenvectors/beta4000_p.csv};
\nextgroupplot[
        ymin=-100,ymax=2200,
            xticklabels={,},
    xtick={},
]
\addplot [
    color=blue,
     very thick,
    solid,smooth
]
table [
    x expr=\thisrow{x}*1000,    
    y expr=\thisrow{T}, 
    col sep=comma
] {figures/Data/flucs_k4016.csv};

\addplot [
    color=orange,
    very thick,
    dashed,smooth
]
table [
    x expr=\thisrowno{0}*1000,    
    y expr=\thisrowno{1}, 
    col sep=comma
] {figures/Data/eigenvectors/beta4000_T.csv};
\nextgroupplot[
        ymin=-0.1,ymax=1.1,
            xticklabels={,},
    xtick={},
]
\addplot [
    color=blue,
     very thick,
    solid,smooth
]
table [
    x expr=\thisrow{x}*1000,    
    y expr=\thisrow{c}, 
    col sep=comma
] {figures/Data/flucs_k4016.csv};

\addplot [
    color=orange,
    very thick,
    dashed,smooth
]
table [
    x expr=\thisrowno{0}*1000,    
    y expr=\thisrowno{1}, 
    col sep=comma
] {figures/Data/eigenvectors/beta4000_c.csv};
\nextgroupplot[
        ymin=-4.5,ymax=4.5,
        ytick={-3,0,3},
            xlabel={$x/\text{mm}$},xlabel style={yshift=5pt}
]
\addplot [
    color=blue,
     very thick,
    solid,smooth
]
table [
    x expr=\thisrow{x}*1000,    
    y expr=\thisrow{RR}/10000, 
    col sep=comma
] {figures/Data/flucs_k4016.csv};

\addplot [
    color=orange,
    very thick,
    dashed,smooth,
]
table [
    x expr=\thisrowno{0}*1000,    
    y expr=\thisrowno{1}/10000, 
    col sep=comma
] {figures/Data/eigenvectors/beta4000_source.csv};
\end{groupplot}
\end{tikzpicture}

%% file: figures/ZirwesSpektrum3D.tex
{\pgfplotsset{compat=1.18,
        label style={font=\footnotesize},
        tick label style={font=\footnotesize}}
\begin{tikzpicture}
\begin{groupplot}[
    group style={
        group name=my plots,
        group size=2 by 1,
        horizontal sep=30pt,  
    },
    width=1.0\linewidth,
    height=4.4cm,
    grid=both,
    xlabel = $x$,
    xmin=0,xmax=4000,
    ymin=-1.7,ymax=1.1,
    legend style={font=\small},
    y label style={yshift=-5pt},
]

\nextgroupplot[
        grid=both,
        xlabel style={yshift=5pt},
        ylabel style={align=center},
        ylabel={$\omega_\text{i}/\text{s}^{-1}$},
        xlabel={$k_y/\text{m}^{-1}=k_z/\text{m}^{-1}$},
        ymin=-1000, ymax=2000,     
        legend columns=2,
            legend style={
                font=\footnotesize,
        at={(0.5,1.02)}, 
        anchor=south,  
inner sep=1pt,
        legend columns=1 
    },
]
        \addplot[dashed, forget plot] coordinates 
        {
            (0,3.099407093782345)
            (15,3.099407093782345)    
        };

        \addplot [
            color=blue,
            fill=none,
            mark=*,
            mark options={fill=white,scale=1,},
            only marks
        ] coordinates 
        {

            (1414,1516)
            (2121,1577)

        };
        \addlegendentry{\gls{DNS}};

    \addplot [
            very thick,
            color=green,
            solid,smooth, forget plot
        ] table     [x expr=1*\thisrow{k}/1.4142,
    y expr=\thisrow{omega_i}, col sep=comma] {figures/Data/ArrheniusDR_2D_9000.csv};
        \label{line:DRReal2DScaled};

        \addplot [
            very thick,
            color=orange,
            very thick, dashed,smooth
        ] table [x=k, y=omega_i, col sep=comma] {figures/Data/ArrheniusDR_3D_beta1_beta2.csv};
        \label{line:DRReal3D}
        \addlegendentry{\gls{LSA}};
        \addplot [
            very thick,
            color=green,
            solid,smooth
        ] coordinates 
        {
            (-2,-1)
            (-2,-1)
        };
        \label{line:DRReal2DScaled}
        \addlegendentry{2D \gls{LSA} scaled in $k$}

%

\end{groupplot}
\end{tikzpicture}}

%% file: figures/DRContourplot.tex
\pgfplotsset{compat=1.18,
        label style={font=\footnotesize},
        tick label style={font=\footnotesize},
        every axis label/.style={font=\footnotesize},}
\begin{tikzpicture}
\begin{axis}[
view={35}{65},
xlabel={$k_y/{\text{m}^{-1}}$},
ylabel={$k_z/{\text{m}^{-1}}$},
zlabel={$\omega_\text{i}/\text{s}^{-1}$},
zmin={0},
zmax={2000},
width={6cm},
height={7cm},
xmin = {0}, xmax={5500},
ymin = {0}, ymax={5500},
smooth,
colorbar horizontal,
    colormap/viridis,
colorbar style={
    height={0.3cm},
    at={(0.5,1.05)},
    anchor=south,
    title={$\omega_\text{i}/\text{s}^{-1}$},
    xmin={0},xmax=1600,
    xtick = {0,500,1000,1600},
}
]
    \addplot3 [surf] table {
1e-05 1e-05 1.0
1e-05 189.65518206896553 183.4399601140508
1e-05 379.31035413793103 403.823385633244
1e-05 568.9655262068966 603.1745132239198
1e-05 758.6206982758621 783.0122486238063
1e-05 948.2758703448276 944.1211754904157
1e-05 1137.9310424137932 1087.0236989590692
1e-05 1327.5862144827586 1212.079230207438
1e-05 1517.2413865517242 1319.5313042543155
1e-05 1706.8965586206898 1409.537758101698
1e-05 1896.5517306896552 1482.192376317345
1e-05 2086.206902758621 1537.541213905814
1e-05 2275.8620748275866 1575.5952384668012
1e-05 2465.517246896552 1596.3402871809953
1e-05 2655.1724189655174 1599.7450039066355
1e-05 2844.8275910344832 1585.7672278769624
1e-05 3034.4827631034486 1554.3591815331822
1e-05 3224.137935172414 1505.4717214394673
1e-05 3413.79310724138 1439.0578573487821
1e-05 3603.4482793103452 1355.0757026115195
1e-05 3793.1034513793106 1253.4909886528246
1e-05 3982.7586234482765 1134.2792541439119
1e-05 4172.413795517241 997.4278041185173
1e-05 4362.068967586207 842.9375238635932
1e-05 4551.724139655173 670.8246267652066
1e-05 4741.379311724138 481.1224138587261
1e-05 4931.034483793103 273.88312618335794
1e-05 5120.689655862069 49.179979522015856
1e-05 5310.344827931034 -1.0
1e-05 5500.0 -1.0

189.65518206896553 1e-05 183.43996011494164
189.65518206896553 189.65518206896553 277.4625922338105
189.65518206896553 379.31035413793103 452.70087381355825
189.65518206896553 568.9655262068966 633.6594869858222
189.65518206896553 758.6206982758621 803.8440965881055
189.65518206896553 948.2758703448276 959.076508516848
189.65518206896553 1137.9310424137932 1098.047021053245
189.65518206896553 1327.5862144827586 1220.2943914999503
189.65518206896553 1517.2413865517242 1325.6429047485087
189.65518206896553 1706.8965586206898 1414.0150367576805
189.65518206896553 1896.5517306896552 1485.362763450685
189.65518206896553 2086.206902758621 1539.6418729354125
189.65518206896553 2275.8620748275866 1576.8032952406793
189.65518206896553 2465.517246896552 1596.791475823738
189.65518206896553 2655.1724189655174 1599.5456547090894
189.65518206896553 2844.8275910344832 1585.002246968315
189.65518206896553 3034.4827631034486 1553.0975217971613
189.65518206896553 3224.137935172414 1503.7702301991935
189.65518206896553 3413.79310724138 1436.9640432809058
189.65518206896553 3603.4482793103452 1352.6297650956863
189.65518206896553 3793.1034513793106 1250.7273325758931
189.65518206896553 3982.7586234482765 1131.227638958411
189.65518206896553 4172.413795517241 994.1142284556413
189.65518206896553 4362.068967586207 839.3849162258066
189.65518206896553 4551.724139655173 667.0533921681504
189.65518206896553 4741.379311724138 477.15087248591567
189.65518206896553 4931.034483793103 269.72787052435115
189.65518206896553 5120.689655862069 44.85616982235024
189.65518206896553 5310.344827931034 -1.0
189.65518206896553 5500.0 -1.0

379.31035413793103 1e-05 403.8233856333286
379.31035413793103 189.65518206896553 452.700873813179
379.31035413793103 379.31035413793103 570.3855949837698
379.31035413793103 568.9655262068966 714.3482765764404
379.31035413793103 758.6206982758621 861.3748984556485
379.31035413793103 948.2758703448276 1001.295316913313
379.31035413793103 1137.9310424137932 1129.5535362507435
379.31035413793103 1327.5862144827586 1243.9443058645004
379.31035413793103 1517.2413865517242 1343.3062257570236
379.31035413793103 1706.8965586206898 1426.9730031594986
379.31035413793103 1896.5517306896552 1494.5270754598105
379.31035413793103 2086.206902758621 1545.6822400394253
379.31035413793103 2275.8620748275866 1580.2251456361746
379.31035413793103 2465.517246896552 1597.985233275639
379.31035413793103 2655.1724189655174 1598.8190843137195
379.31035413793103 2844.8275910344832 1582.6023328293936
379.31035413793103 3034.4827631034486 1549.2256561428674
379.31035413793103 3224.137935172414 1498.5930039408074
379.31035413793103 3413.79310724138 1430.621069662512
379.31035413793103 3603.4482793103452 1345.2394566580185
379.31035413793103 3793.1034513793106 1242.3912380158372
379.31035413793103 3982.7586234482765 1122.0337486679368
379.31035413793103 4172.413795517241 984.1395299053397
379.31035413793103 4362.068967586207 828.6973962887305
379.31035413793103 4551.724139655173 655.7136269926737
379.31035413793103 4741.379311724138 465.2133069823517
379.31035413793103 4931.034483793103 257.2418628037294
379.31035413793103 5120.689655862069 31.866857459105177
379.31035413793103 5310.344827931034 -1.0
379.31035413793103 5500.0 -1.0

568.9655262068966 1e-05 603.1745132245446
568.9655262068966 189.65518206896553 633.6594869876428
568.9655262068966 379.31035413793103 714.348276577256
568.9655262068966 568.9655262068966 823.8014847952991
568.9655262068966 758.6206982758621 944.1211786945198
568.9655262068966 948.2758703448276 1064.1279798781534
568.9655262068966 1137.9310424137932 1177.4163934866624
568.9655262068966 1327.5862144827586 1280.3161211287218
568.9655262068966 1517.2413865517242 1370.6517045890814
568.9655262068966 1706.8965586206898 1447.072092101449
568.9655262068966 1896.5517306896552 1508.6938851180587
568.9655262068966 2086.206902758621 1554.9082905949076
568.9655262068966 2275.8620748275866 1585.2739212194904
568.9655262068966 2465.517246896552 1599.4556852015485
568.9655262068966 2655.1724189655174 1597.1891341159844
568.9655262068966 2844.8275910344832 1578.2592738640494
568.9655262068966 3034.4827631034486 1542.487808054937
568.9655262068966 3224.137935172414 1489.7254141251196
568.9655262068966 3413.79310724138 1419.847088217401
568.9655262068966 3603.4482793103452 1332.749401318704
568.9655262068966 3793.1034513793106 1228.348973971014
568.9655262068966 3982.7586234482765 1106.581752592727
568.9655262068966 4172.413795517241 967.4028376088454
568.9655262068966 4362.068967586207 810.7867185376911
568.9655262068966 4551.724139655173 636.7278395076942
568.9655262068966 4741.379311724138 445.24146601319353
568.9655262068966 4931.034483793103 236.36485934805205
568.9655262068966 5120.689655862069 10.15879640746516
568.9655262068966 5310.344827931034 -1.0
568.9655262068966 5500.0 -1.0

758.6206982758621 1e-05 783.0122486239927
758.6206982758621 189.65518206896553 803.8440965871421
758.6206982758621 379.31035413793103 861.3748984557067
758.6206982758621 568.9655262068966 944.1211786931112
758.6206982758621 758.6206982758621 1040.013988539792
758.6206982758621 948.2758703448276 1139.5678157904626
758.6206982758621 1137.9310424137932 1236.2220339555192
758.6206982758621 1327.5862144827586 1325.6429060250018
758.6206982758621 1517.2413865517242 1404.9790472248967
758.6206982758621 1706.8965586206898 1472.3220289073763
758.6206982758621 1896.5517306896552 1526.3590720694528
758.6206982758621 2086.206902758621 1566.1578349768633
758.6206982758621 2275.8620748275866 1591.0339457406471
758.6206982758621 2465.517246896552 1600.4689160880957
758.6206982758621 2655.1724189655174 1594.05865613808
758.6206982758621 2844.8275910344832 1571.4807412721307
758.6206982758621 3034.4827631034486 1532.4733291697642
758.6206982758621 3224.137935172414 1476.8214298703178
758.6206982758621 3413.79310724138 1404.3478949465014
758.6206982758621 3603.4482793103452 1314.9074895392237
758.6206982758621 3793.1034513793106 1208.3830184251487
758.6206982758621 3982.7586234482765 1084.6828537110896
758.6206982758621 4172.413795517241 943.7394493557699
758.6206982758621 4362.068967586207 785.5085813006269
758.6206982758621 4551.724139655173 609.9691541319164
758.6206982758621 4741.379311724138 417.12348597139726
758.6206982758621 4931.034483793103 206.99803591844693
758.6206982758621 5120.689655862069 -1.0
758.6206982758621 5310.344827931034 -1.0
758.6206982758621 5500.0 -1.0

948.2758703448276 1e-05 944.1211754905075
948.2758703448276 189.65518206896553 959.0765085194423
948.2758703448276 379.31035413793103 1001.2953169136115
948.2758703448276 568.9655262068966 1064.127979877862
948.2758703448276 758.6206982758621 1139.567815790507
948.2758703448276 948.2758703448276 1220.2943932143974
948.2758703448276 1137.9310424137932 1300.4871454002325
948.2758703448276 1327.5862144827586 1375.826617123675
948.2758703448276 1517.2413865517242 1443.1960140469737
948.2758703448276 1706.8965586206898 1500.3559104245044
948.2758703448276 1896.5517306896552 1545.6822404502418
948.2758703448276 2086.206902758621 1577.977402367435
948.2758703448276 2275.8620748275866 1596.3402873784207
948.2758703448276 2465.517246896552 1600.0784666794643
948.2758703448276 2655.1724189655174 1588.6491109794329
948.2758703448276 2844.8275910344832 1561.6191276993718
948.2758703448276 3034.4827631034486 1518.6380979529101
948.2758703448276 3224.137935172414 1459.4197690905546
948.2758703448276 3413.79310724138 1383.7293121809782
948.2758703448276 3603.4482793103452 1291.3745078478867
948.2758703448276 3793.1034513793106 1182.199647201721
948.2758703448276 3982.7586234482765 1056.0813430626529
948.2758703448276 4172.413795517241 912.9257174144439
948.2758703448276 4362.068967586207 752.6666122871154
948.2758703448276 4551.724139655173 575.2645959916115
948.2758703448276 4741.379311724138 380.70662463868166
948.2758703448276 4931.034483793103 169.00628475458643
948.2758703448276 5120.689655862069 -1.0
948.2758703448276 5310.344827931034 -1.0
948.2758703448276 5500.0 -1.0

1137.9310424137932 1e-05 1087.0236989579532
1137.9310424137932 189.65518206896553 1098.047021052426
1137.9310424137932 379.31035413793103 1129.55353625126
1137.9310424137932 568.9655262068966 1177.4163934861776
1137.9310424137932 758.6206982758621 1236.2220339540165
1137.9310424137932 948.2758703448276 1300.487145400117
1137.9310424137932 1137.9310424137932 1365.3814868993702
1137.9310424137932 1327.5862144827586 1426.9730040418885
1137.9310424137932 1517.2413865517242 1482.1923776659512
1137.9310424137932 1706.8965586206898 1528.6879313505165
1137.9310424137932 1896.5517306896552 1564.664948138694
1137.9310424137932 2086.206902758621 1588.7468936096511
1137.9310424137932 2275.8620748275866 1599.867230745815
1137.9310424137932 2465.517246896552 1597.189134031319
1137.9310424137932 2655.1724189655174 1580.0473199054213
1137.9310424137932 2844.8275910344832 1547.9063455476867
1137.9310424137932 3034.4827631034486 1500.3308096305486
1137.9310424137932 3224.137935172414 1436.9640423960097
1137.9310424137932 3413.79310724138 1357.5128294947676
1137.9310424137932 3603.4482793103452 1261.7364357134902
1137.9310424137932 3793.1034513793106 1149.438715620474
1137.9310424137932 3982.7586234482765 1020.4624670590331
1137.9310424137932 4172.413795517241 874.6854426288999
1137.9310424137932 4362.068967586207 712.0176172513509
1137.9310424137932 4551.724139655173 532.3994407429232
1137.9310424137932 4741.379311724138 335.80089967715867
1137.9310424137932 4931.034483793103 122.22128600320639
1137.9310424137932 5120.689655862069 -1.0
1137.9310424137932 5310.344827931034 -1.0
1137.9310424137932 5500.0 -1.0

1327.5862144827586 1e-05 1212.0792302060825
1327.5862144827586 189.65518206896553 1220.2943915002347
1327.5862144827586 379.31035413793103 1243.944305864072
1327.5862144827586 568.9655262068966 1280.3161211290417
1327.5862144827586 758.6206982758621 1325.6429060254284
1327.5862144827586 948.2758703448276 1375.8266171241319
1327.5862144827586 1137.9310424137932 1426.9730040421534
1327.5862144827586 1327.5862144827586 1475.6729252888979
1327.5862144827586 1517.2413865517242 1519.0855447563285
1327.5862144827586 1706.8965586206898 1554.908290963028
1327.5862144827586 1896.5517306896552 1581.2994579056622
1327.5862144827586 2086.206902758621 1596.791476010742
1327.5862144827586 2275.8620748275866 1600.2122482975867
1327.5862144827586 2465.517246896552 1590.620238822547
1327.5862144827586 2655.1724189655174 1567.2534847435868
1327.5862144827586 2844.8275910344832 1529.4906076125496
1327.5862144827586 3034.4827631034486 1476.8214294720792
1327.5862144827586 3224.137935172414 1408.8249906186627
1327.5862144827586 3413.79310724138 1325.1531616454604
1327.5862144827586 3603.4482793103452 1225.5184481052552
1327.5862144827586 3793.1034513793106 1109.6849349065897
1327.5862144827586 3982.7586234482765 977.4615944369837
1327.5862144827586 4172.413795517241 828.6973947784104
1327.5862144827586 4362.068967586207 663.2778045030657
1327.5862144827586 4551.724139655173 481.12241134861824
1327.5862144827586 4741.379311724138 282.1834646093557
1327.5862144827586 4931.034483793103 66.44522435061072
1327.5862144827586 5120.689655862069 -1.0
1327.5862144827586 5310.344827931034 -1.0
1327.5862144827586 5500.0 -1.0

1517.2413865517242 1e-05 1319.5313042542698
1517.2413865517242 189.65518206896553 1325.6429047477172
1517.2413865517242 379.31035413793103 1343.306225756324
1517.2413865517242 568.9655262068966 1370.6517045887592
1517.2413865517242 758.6206982758621 1404.9790472255104
1517.2413865517242 948.2758703448276 1443.1960140470453
1517.2413865517242 1137.9310424137932 1482.192377666102
1517.2413865517242 1327.5862144827586 1519.0855447549939
1517.2413865517242 1517.2413865517242 1551.341081707312
1517.2413865517242 1706.8965586206898 1576.8032957441767
1517.2413865517242 1896.5517306896552 1593.6741238797479
1517.2413865517242 2086.206902758621 1600.4689161029078
1517.2413865517242 2275.8620748275866 1595.9664125374559
1517.2413865517242 2465.517246896552 1579.1616954425408
1517.2413865517242 2655.1724189655174 1549.2256555911936
1517.2413865517242 2844.8275910344832 1505.4717203648706
1517.2413865517242 3034.4827631034486 1447.3293070471805
1517.2413865517242 3224.137935172414 1374.3230012303745
1517.2413865517242 3413.79310724138 1286.0563974634354
1517.2413865517242 3603.4482793103452 1182.1996465792245
1517.2413865517242 3793.1034513793106 1062.47991655183
1517.2413865517242 3982.7586234482765 926.674137457323
1517.2413865517242 4172.413795517241 774.6035464872778
1517.2413865517242 4362.068967586207 606.1296701166957
1517.2413865517242 4551.724139655173 421.15147935388586
1517.2413865517242 4741.379311724138 219.6035346404069
1517.2413865517242 4931.034483793103 1.4550045094874804
1517.2413865517242 5120.689655862069 -1.0
1517.2413865517242 5310.344827931034 -1.0
1517.2413865517242 5500.0 -1.0

1706.8965586206898 1e-05 1409.537758101382
1706.8965586206898 189.65518206896553 1414.0150367588926
1706.8965586206898 379.31035413793103 1426.97300315946
1706.8965586206898 568.9655262068966 1447.0720921037987
1706.8965586206898 758.6206982758621 1472.3220289077244
1706.8965586206898 948.2758703448276 1500.3559104247583
1706.8965586206898 1137.9310424137932 1528.6879313510863
1706.8965586206898 1327.5862144827586 1554.9082909630586
1706.8965586206898 1517.2413865517242 1576.8032957436876
1706.8965586206898 1706.8965586206898 1592.4115968903216
1706.8965586206898 1896.5517306896552 1600.0359879700527
1706.8965586206898 2086.206902758621 1598.2291845242835
1706.8965586206898 2275.8620748275866 1585.7672273966919
1706.8965586206898 2465.517246896552 1561.6191274527769
1706.8965586206898 2655.1724189655174 1524.9174972054107
1706.8965586206898 2844.8275910344832 1474.9323786212033
1706.8965586206898 3034.4827631034486 1411.048997235472
1706.8965586206898 3224.137935172414 1332.7494002572043
1706.8965586206898 3413.79310724138 1239.5975883521096
1706.8965586206898 3603.4482793103452 1131.2276370211137
1706.8965586206898 3793.1034513793106 1007.33430769405
1706.8965586206898 3982.7586234482765 867.6657056296056
1706.8965586206898 4172.413795517241 712.0176164668815
1706.8965586206898 4362.068967586207 540.2292293270621
1706.8965586206898 4551.724139655173 352.18002469720113
1706.8965586206898 4741.379311724138 147.78766877252383
1706.8965586206898 4931.034483793103 -1.0
1706.8965586206898 5120.689655862069 -1.0
1706.8965586206898 5310.344827931034 -1.0
1706.8965586206898 5500.0 -1.0

1896.5517306896552 1e-05 1482.1923763162094
1896.5517306896552 189.65518206896553 1485.362763452036
1896.5517306896552 379.31035413793103 1494.5270754598591
1896.5517306896552 568.9655262068966 1508.693885118265
1896.5517306896552 758.6206982758621 1526.3590720694724
1896.5517306896552 948.2758703448276 1545.6822404504078
1896.5517306896552 1137.9310424137932 1564.6649481393154
1896.5517306896552 1327.5862144827586 1581.299457904307
1896.5517306896552 1517.2413865517242 1593.6741238790114
1896.5517306896552 1706.8965586206898 1600.0359879700036
1896.5517306896552 1896.5517306896552 1598.8190841974051
1896.5517306896552 2086.206902758621 1588.6491108353935
1896.5517306896552 2275.8620748275866 1568.3339328588718
1896.5517306896552 2465.517246896552 1536.8469468624596
1896.5517306896552 2655.1724189655174 1493.3079234675047
1896.5517306896552 2844.8275910344832 1436.9640419535517
1896.5517306896552 3034.4827631034486 1367.17252339926
1896.5517306896552 3224.137935172414 1283.385451047493
1896.5517306896552 3413.79310724138 1185.1368983639777
1896.5517306896552 3603.4482793103452 1072.0322439640875
1896.5517306896552 3793.1034513793106 943.739447921698
1896.5517306896552 3982.7586234482765 799.9820368891098
1896.5517306896552 4172.413795517241 640.5335592268593
1896.5517306896552 4362.068967586207 465.2133044626196
1896.5517306896552 4551.724139655173 273.88312268184063
1896.5517306896552 4741.379311724138 66.445223440066
1896.5517306896552 4931.034483793103 -1.0
1896.5517306896552 5120.689655862069 -1.0
1896.5517306896552 5310.344827931034 -1.0
1896.5517306896552 5500.0 -1.0

2086.206902758621 1e-05 1537.5412139067648
2086.206902758621 189.65518206896553 1539.6418729357083
2086.206902758621 379.31035413793103 1545.682240040252
2086.206902758621 568.9655262068966 1554.9082905954074
2086.206902758621 758.6206982758621 1566.1578349765857
2086.206902758621 948.2758703448276 1577.9774023676166
2086.206902758621 1137.9310424137932 1588.746893609767
2086.206902758621 1327.5862144827586 1596.7914760122649
2086.206902758621 1517.2413865517242 1600.4689161032575
2086.206902758621 1706.8965586206898 1598.229184523459
2086.206902758621 1896.5517306896552 1588.649110835012
2086.206902758621 2086.206902758621 1570.4476568383884
2086.206902758621 2275.8620748275866 1542.4878074703138
2086.206902758621 2465.517246896552 1503.7702291152773
2086.206902758621 2655.1724189655174 1453.422558792378
2086.206902758621 2844.8275910344832 1390.6869475571757
2086.206902758621 3034.4827631034486 1314.9074884497436
2086.206902758621 3224.137935172414 1225.5184475052292
2086.206902758621 3413.79310724138 1122.0337467243405
2086.206902758621 3603.4482793103452 1004.0378598180787
2086.206902758621 3793.1034513793106 871.1781185630127
2086.206902758621 3982.7586234482765 723.1583461199905
2086.206902758621 4172.413795517241 559.7337021198816
2086.206902758621 4362.068967586207 380.70662292640054
2086.206902758621 4551.724139655173 185.92375592192366
2086.206902758621 4741.379311724138 -1.0
2086.206902758621 4931.034483793103 -1.0
2086.206902758621 5120.689655862069 -1.0
2086.206902758621 5310.344827931034 -1.0
2086.206902758621 5500.0 -1.0

2275.8620748275866 1e-05 1575.5952384666964
2275.8620748275866 189.65518206896553 1576.8032952423473
2275.8620748275866 379.31035413793103 1580.2251456351523
2275.8620748275866 568.9655262068966 1585.2739212202066
2275.8620748275866 758.6206982758621 1591.0339457405662
2275.8620748275866 948.2758703448276 1596.3402873764562
2275.8620748275866 1137.9310424137932 1599.867230746114
2275.8620748275866 1327.5862144827586 1600.2122482969553
2275.8620748275866 1517.2413865517242 1595.966412537064
2275.8620748275866 1706.8965586206898 1585.7672273983035
2275.8620748275866 1896.5517306896552 1568.3339328595002
2275.8620748275866 2086.206902758621 1542.4878074708554
2275.8620748275866 2275.8620748275866 1507.160982790887
2275.8620748275866 2465.517246896552 1461.3972623078039
2275.8620748275866 2655.1724189655174 1404.347893999229
2275.8620748275866 2844.8275910344832 1335.2645363095248
2275.8620748275866 3034.4827631034486 1253.4909863268997
2275.8620748275866 3224.137935172414 1158.4546919461486
2275.8620748275866 3413.79310724138 1049.6586654326409
2275.8620748275866 3603.4482793103452 926.674136736435
2275.8620748275866 3793.1034513793106 789.1341048813665
2275.8620748275866 3982.7586234482765 636.7278370975496
2275.8620748275866 4172.413795517241 469.19630646840756
2275.8620748275866 4362.068967586207 286.3285328554771
2275.8620748275866 4551.724139655173 87.95878795380668
2275.8620748275866 4741.379311724138 -1.0
2275.8620748275866 4931.034483793103 -1.0
2275.8620748275866 5120.689655862069 -1.0
2275.8620748275866 5310.344827931034 -1.0
2275.8620748275866 5500.0 -1.0

2465.517246896552 1e-05 1596.3402871806811
2465.517246896552 189.65518206896553 1596.7914758252523
2465.517246896552 379.31035413793103 1597.9852332756536
2465.517246896552 568.9655262068966 1599.4556852022147
2465.517246896552 758.6206982758621 1600.4689160888083
2465.517246896552 948.2758703448276 1600.078466679649
2465.517246896552 1137.9310424137932 1597.1891340315456
2465.517246896552 1327.5862144827586 1590.6202388223614
2465.517246896552 1517.2413865517242 1579.161695442911
2465.517246896552 1706.8965586206898 1561.6191274527919
2465.517246896552 1896.5517306896552 1536.8469468630803
2465.517246896552 2086.206902758621 1503.7702291145977
2465.517246896552 2275.8620748275866 1461.397262308196
2465.517246896552 2465.517246896552 1408.8249901473673
2465.517246896552 2655.1724189655174 1345.239455089063
2465.517246896552 2844.8275910344832 1269.9130081267472
2465.517246896552 3034.4827631034486 1182.1996459609368
2465.517246896552 3224.137935172414 1081.5294559677436
2465.517246896552 3413.79310724138 967.4028354798365
2465.517246896552 3603.4482793103452 839.3849132278963
2465.517246896552 3793.1034513793106 697.1004308391457
2465.517246896552 3982.7586234482765 540.2292285037895
2465.517246896552 4172.413795517241 368.5024074137277
2465.517246896552 4362.068967586207 181.699202078275
2465.517246896552 4551.724139655173 -1.0
2465.517246896552 4741.379311724138 -1.0
2465.517246896552 4931.034483793103 -1.0
2465.517246896552 5120.689655862069 -1.0
2465.517246896552 5310.344827931034 -1.0
2465.517246896552 5500.0 -1.0

2655.1724189655174 1e-05 1599.7450039063217
2655.1724189655174 189.65518206896553 1599.545654709209
2655.1724189655174 379.31035413793103 1598.8190843135465
2655.1724189655174 568.9655262068966 1597.189134115527
2655.1724189655174 758.6206982758621 1594.0586561390926
2655.1724189655174 948.2758703448276 1588.6491109794365
2655.1724189655174 1137.9310424137932 1580.0473199042558
2655.1724189655174 1327.5862144827586 1567.2534847430325
2655.1724189655174 1517.2413865517242 1549.2256555918282
2655.1724189655174 1706.8965586206898 1524.9174972060514
2655.1724189655174 1896.5517306896552 1493.3079234671611
2655.1724189655174 2086.206902758621 1453.422558792371
2655.1724189655174 2275.8620748275866 1404.3478939986971
2655.1724189655174 2465.517246896552 1345.2394550886174
2655.1724189655174 2655.1724189655174 1275.325408674515
2655.1724189655174 2844.8275910344832 1193.9069187048349
2655.1724189655174 3034.4827631034486 1100.3563564573592
2655.1724189655174 3224.137935172414 994.1142256597946
2655.1724189655174 3413.79310724138 874.6854411534835
2655.1724189655174 3603.4482793103452 741.6354129758336
2655.1724189655174 3793.1034513793106 594.5862430777368
2655.1724189655174 3982.7586234482765 433.2132367828144
2655.1724189655174 4172.413795517241 257.24185928979387
2655.1724189655174 4362.068967586207 66.44522252884917
2655.1724189655174 4551.724139655173 -1.0
2655.1724189655174 4741.379311724138 -1.0
2655.1724189655174 4931.034483793103 -1.0
2655.1724189655174 5120.689655862069 -1.0
2655.1724189655174 5310.344827931034 -1.0
2655.1724189655174 5500.0 -1.0

2844.8275910344832 1e-05 1585.767227876377
2844.8275910344832 189.65518206896553 1585.002246966465
2844.8275910344832 379.31035413793103 1582.6023328300766
2844.8275910344832 568.9655262068966 1578.2592738627059
2844.8275910344832 758.6206982758621 1571.4807412726188
2844.8275910344832 948.2758703448276 1561.6191276997902
2844.8275910344832 1137.9310424137932 1547.9063455487305
2844.8275910344832 1327.5862144827586 1529.490607611667
2844.8275910344832 1517.2413865517242 1505.4717203649084
2844.8275910344832 1706.8965586206898 1474.9323786210023
2844.8275910344832 1896.5517306896552 1436.964041953879
2844.8275910344832 2086.206902758621 1390.6869475559606
2844.8275910344832 2275.8620748275866 1335.2645363088918
2844.8275910344832 2465.517246896552 1269.9130081275987
2844.8275910344832 2655.1724189655174 1193.9069187054952
2844.8275910344832 2844.8275910344832 1106.581750626537
2844.8275910344832 3034.4827631034486 1007.3343069974517
2844.8275910344832 3224.137935172414 895.6216414915336
2844.8275910344832 3413.79310724138 770.959092684008
2844.8275910344832 3603.4482793103452 632.9178555906262
2844.8275910344832 3793.1034513793106 481.1224096745259
2844.8275910344832 3982.7586234482765 315.2480340613006
2844.8275910344832 4172.413795517241 135.0185756944852
2844.8275910344832 4362.068967586207 -1.0
2844.8275910344832 4551.724139655173 -1.0
2844.8275910344832 4741.379311724138 -1.0
2844.8275910344832 4931.034483793103 -1.0
2844.8275910344832 5120.689655862069 -1.0
2844.8275910344832 5310.344827931034 -1.0
2844.8275910344832 5500.0 -1.0

3034.4827631034486 1e-05 1554.3591815332584
3034.4827631034486 189.65518206896553 1553.097521797515
3034.4827631034486 379.31035413793103 1549.2256561425147
3034.4827631034486 568.9655262068966 1542.4878080553458
3034.4827631034486 758.6206982758621 1532.4733291697084
3034.4827631034486 948.2758703448276 1518.6380979525325
3034.4827631034486 1137.9310424137932 1500.3308096289934
3034.4827631034486 1327.5862144827586 1476.8214294726627
3034.4827631034486 1517.2413865517242 1447.3293070468392
3034.4827631034486 1706.8965586206898 1411.0489972355647
3034.4827631034486 1896.5517306896552 1367.1725233989887
3034.4827631034486 2086.206902758621 1314.9074884484266
3034.4827631034486 2275.8620748275866 1253.49098632642
3034.4827631034486 2465.517246896552 1182.1996459611169
3034.4827631034486 2655.1724189655174 1100.356356456568
3034.4827631034486 2844.8275910344832 1007.3343069977043
3034.4827631034486 3034.4827631034486 902.5589679710297
3034.4827631034486 3224.137935172414 785.5085790027304
3034.4827631034486 3413.79310724138 655.7136237970146
3034.4827631034486 3603.4482793103452 512.7556813702295
3034.4827631034486 3793.1034513793106 356.26596055944947
3034.4827631034486 3982.7586234482765 185.92375503286667
3034.4827631034486 4172.413795517241 1.4550026738165798
3034.4827631034486 4362.068967586207 -1.0
3034.4827631034486 4551.724139655173 -1.0
3034.4827631034486 4741.379311724138 -1.0
3034.4827631034486 4931.034483793103 -1.0
3034.4827631034486 5120.689655862069 -1.0
3034.4827631034486 5310.344827931034 -1.0
3034.4827631034486 5500.0 -1.0

3224.137935172414 1e-05 1505.47172143956
3224.137935172414 189.65518206896553 1503.7702301982927
3224.137935172414 379.31035413793103 1498.5930039412665
3224.137935172414 568.9655262068966 1489.7254141250169
3224.137935172414 758.6206982758621 1476.821429870587
3224.137935172414 948.2758703448276 1459.41976909036
3224.137935172414 1137.9310424137932 1436.9640423955336
3224.137935172414 1327.5862144827586 1408.824990619772
3224.137935172414 1517.2413865517242 1374.3230012307579
3224.137935172414 1706.8965586206898 1332.7494002571027
3224.137935172414 1896.5517306896552 1283.3854510471815
3224.137935172414 2086.206902758621 1225.5184475065603
3224.137935172414 2275.8620748275866 1158.4546919465276
3224.137935172414 2465.517246896552 1081.5294559683884
3224.137935172414 2655.1724189655174 994.1142256607573
3224.137935172414 2844.8275910344832 895.6216414915236
3224.137935172414 3034.4827631034486 785.5085790030298
3224.137935172414 3224.137935172414 663.2778029079834
3224.137935172414 3413.79310724138 528.4785850448325
3224.137935172414 3603.4482793103452 380.7066220677252
3224.137935172414 3793.1034513793106 219.6035328729772
3224.137935172414 3982.7586234482765 44.85616526155809
3224.137935172414 4172.413795517241 -1.0
3224.137935172414 4362.068967586207 -1.0
3224.137935172414 4551.724139655173 -1.0
3224.137935172414 4741.379311724138 -1.0
3224.137935172414 4931.034483793103 -1.0
3224.137935172414 5120.689655862069 -1.0
3224.137935172414 5310.344827931034 -1.0
3224.137935172414 5500.0 -1.0

3413.79310724138 1e-05 1439.0578573494315
3413.79310724138 189.65518206896553 1436.9640432801596
3413.79310724138 379.31035413793103 1430.6210696628086
3413.79310724138 568.9655262068966 1419.8470882169986
3413.79310724138 758.6206982758621 1404.3478949472358
3413.79310724138 948.2758703448276 1383.7293121804516
3413.79310724138 1137.9310424137932 1357.5128294931083
3413.79310724138 1327.5862144827586 1325.1531616445568
3413.79310724138 1517.2413865517242 1286.0563974627887
3413.79310724138 1706.8965586206898 1239.5975883522003
3413.79310724138 1896.5517306896552 1185.136898363874
3413.79310724138 2086.206902758621 1122.0337467228003
3413.79310724138 2275.8620748275866 1049.658665432987
3413.79310724138 2465.517246896552 967.4028354798234
3413.79310724138 2655.1724189655174 874.6854411526933
3413.79310724138 2844.8275910344832 770.9590926828523
3413.79310724138 3034.4827631034486 655.7136237970224
3413.79310724138 3224.137935172414 528.4785850451644
3413.79310724138 3413.79310724138 388.82474155776026
3413.79310724138 3603.4482793103452 236.36485582047726
3413.79310724138 3793.1034513793106 70.75400216881644
3413.79310724138 3982.7586234482765 -1.0
3413.79310724138 4172.413795517241 -1.0
3413.79310724138 4362.068967586207 -1.0
3413.79310724138 4551.724139655173 -1.0
3413.79310724138 4741.379311724138 -1.0
3413.79310724138 4931.034483793103 -1.0
3413.79310724138 5120.689655862069 -1.0
3413.79310724138 5310.344827931034 -1.0
3413.79310724138 5500.0 -1.0

3603.4482793103452 1e-05 1355.0757026116232
3603.4482793103452 189.65518206896553 1352.6297650979443
3603.4482793103452 379.31035413793103 1345.2394566584392
3603.4482793103452 568.9655262068966 1332.7494013182497
3603.4482793103452 758.6206982758621 1314.9074895392405
3603.4482793103452 948.2758703448276 1291.374507847859
3603.4482793103452 1137.9310424137932 1261.7364357135589
3603.4482793103452 1327.5862144827586 1225.5184481056065
3603.4482793103452 1517.2413865517242 1182.1996465780844
3603.4482793103452 1706.8965586206898 1131.2276370210923
3603.4482793103452 1896.5517306896552 1072.0322439632057
3603.4482793103452 2086.206902758621 1004.0378598182374
3603.4482793103452 2275.8620748275866 926.6741367366963
3603.4482793103452 2465.517246896552 839.3849132263283
3603.4482793103452 2655.1724189655174 741.6354129763652
3603.4482793103452 2844.8275910344832 632.9178555906972
3603.4482793103452 3034.4827631034486 512.7556813707379
3603.4482793103452 3224.137935172414 380.7066220661809
3603.4482793103452 3413.79310724138 236.36485582055275
3603.4482793103452 3603.4482793103452 79.36247566129077
3603.4482793103452 3793.1034513793106 -1.0
3603.4482793103452 3982.7586234482765 -1.0
3603.4482793103452 4172.413795517241 -1.0
3603.4482793103452 4362.068967586207 -1.0
3603.4482793103452 4551.724139655173 -1.0
3603.4482793103452 4741.379311724138 -1.0
3603.4482793103452 4931.034483793103 -1.0
3603.4482793103452 5120.689655862069 -1.0
3603.4482793103452 5310.344827931034 -1.0
3603.4482793103452 5500.0 -1.0

3793.1034513793106 1e-05 1253.4909886525766
3793.1034513793106 189.65518206896553 1250.7273325772076
3793.1034513793106 379.31035413793103 1242.39123801442
3793.1034513793106 568.9655262068966 1228.348973968262
3793.1034513793106 758.6206982758621 1208.3830184245755
3793.1034513793106 948.2758703448276 1182.1996472012522
3793.1034513793106 1137.9310424137932 1149.4387156192774
3793.1034513793106 1327.5862144827586 1109.6849349085735
3793.1034513793106 1517.2413865517242 1062.4799165512115
3793.1034513793106 1706.8965586206898 1007.3343076935471
3793.1034513793106 1896.5517306896552 943.7394479229347
3793.1034513793106 2086.206902758621 871.1781185635891
3793.1034513793106 2275.8620748275866 789.134104881173
3793.1034513793106 2465.517246896552 697.1004308397537
3793.1034513793106 2655.1724189655174 594.5862430801385
3793.1034513793106 2844.8275910344832 481.12240967455386
3793.1034513793106 3034.4827631034486 356.2659605592762
3793.1034513793106 3224.137935172414 219.60353287295084
3793.1034513793106 3413.79310724138 70.75400216987828
3793.1034513793106 3603.4482793103452 -1.0
3793.1034513793106 3793.1034513793106 -1.0
3793.1034513793106 3982.7586234482765 -1.0
3793.1034513793106 4172.413795517241 -1.0
3793.1034513793106 4362.068967586207 -1.0
3793.1034513793106 4551.724139655173 -1.0
3793.1034513793106 4741.379311724138 -1.0
3793.1034513793106 4931.034483793103 -1.0
3793.1034513793106 5120.689655862069 -1.0
3793.1034513793106 5310.344827931034 -1.0
3793.1034513793106 5500.0 -1.0

3982.7586234482765 1e-05 1134.2792541439446
3982.7586234482765 189.65518206896553 1131.2276389575763
3982.7586234482765 379.31035413793103 1122.0337486688852
3982.7586234482765 568.9655262068966 1106.581752592631
3982.7586234482765 758.6206982758621 1084.6828537096462
3982.7586234482765 948.2758703448276 1056.081343062863
3982.7586234482765 1137.9310424137932 1020.4624670602602
3982.7586234482765 1327.5862144827586 977.4615944367558
3982.7586234482765 1517.2413865517242 926.6741374576611
3982.7586234482765 1706.8965586206898 867.6657056302336
3982.7586234482765 1896.5517306896552 799.9820368892545
3982.7586234482765 2086.206902758621 723.1583461197342
3982.7586234482765 2275.8620748275866 636.7278370968827
3982.7586234482765 2465.517246896552 540.2292285035896
3982.7586234482765 2655.1724189655174 433.21323678323597
3982.7586234482765 2844.8275910344832 315.248034062
3982.7586234482765 3034.4827631034486 185.923755031979
3982.7586234482765 3224.137935172414 44.85616526167178
3982.7586234482765 3413.79310724138 -1.0
3982.7586234482765 3603.4482793103452 -1.0
3982.7586234482765 3793.1034513793106 -1.0
3982.7586234482765 3982.7586234482765 -1.0
3982.7586234482765 4172.413795517241 -1.0
3982.7586234482765 4362.068967586207 -1.0
3982.7586234482765 4551.724139655173 -1.0
3982.7586234482765 4741.379311724138 -1.0
3982.7586234482765 4931.034483793103 -1.0
3982.7586234482765 5120.689655862069 -1.0
3982.7586234482765 5310.344827931034 -1.0
3982.7586234482765 5500.0 -1.0

4172.413795517241 1e-05 997.4278041187827
4172.413795517241 189.65518206896553 994.1142284561047
4172.413795517241 379.31035413793103 984.1395299050373
4172.413795517241 568.9655262068966 967.4028376089559
4172.413795517241 758.6206982758621 943.7394493556444
4172.413795517241 948.2758703448276 912.9257174146385
4172.413795517241 1137.9310424137932 874.685442628662
4172.413795517241 1327.5862144827586 828.697394778884
4172.413795517241 1517.2413865517242 774.6035464870868
4172.413795517241 1706.8965586206898 712.0176164663283
4172.413795517241 1896.5517306896552 640.5335592270435
4172.413795517241 2086.206902758621 559.7337021199821
4172.413795517241 2275.8620748275866 469.1963064690085
4172.413795517241 2465.517246896552 368.5024074129642
4172.413795517241 2655.1724189655174 257.2418592901595
4172.413795517241 2844.8275910344832 135.01857569508184
4172.413795517241 3034.4827631034486 1.4550026734414132
4172.413795517241 3224.137935172414 -1.0
4172.413795517241 3413.79310724138 -1.0
4172.413795517241 3603.4482793103452 -1.0
4172.413795517241 3793.1034513793106 -1.0
4172.413795517241 3982.7586234482765 -1.0
4172.413795517241 4172.413795517241 -1.0
4172.413795517241 4362.068967586207 -1.0
4172.413795517241 4551.724139655173 -1.0
4172.413795517241 4741.379311724138 -1.0
4172.413795517241 4931.034483793103 -1.0
4172.413795517241 5120.689655862069 -1.0
4172.413795517241 5310.344827931034 -1.0
4172.413795517241 5500.0 -1.0

4362.068967586207 1e-05 842.9375238635635
4362.068967586207 189.65518206896553 839.3849162266802
4362.068967586207 379.31035413793103 828.6973962883133
4362.068967586207 568.9655262068966 810.7867185377127
4362.068967586207 758.6206982758621 785.5085813001999
4362.068967586207 948.2758703448276 752.6666122867884
4362.068967586207 1137.9310424137932 712.0176172508541
4362.068967586207 1327.5862144827586 663.2778045027533
4362.068967586207 1517.2413865517242 606.1296701152048
4362.068967586207 1706.8965586206898 540.2292293260537
4362.068967586207 1896.5517306896552 465.21330446291154
4362.068967586207 2086.206902758621 380.7066229257639
4362.068967586207 2275.8620748275866 286.3285328548973
4362.068967586207 2465.517246896552 181.6992020779162
4362.068967586207 2655.1724189655174 66.44522252968818
4362.068967586207 2844.8275910344832 -1.0
4362.068967586207 3034.4827631034486 -1.0
4362.068967586207 3224.137935172414 -1.0
4362.068967586207 3413.79310724138 -1.0
4362.068967586207 3603.4482793103452 -1.0
4362.068967586207 3793.1034513793106 -1.0
4362.068967586207 3982.7586234482765 -1.0
4362.068967586207 4172.413795517241 -1.0
4362.068967586207 4362.068967586207 -1.0
4362.068967586207 4551.724139655173 -1.0
4362.068967586207 4741.379311724138 -1.0
4362.068967586207 4931.034483793103 -1.0
4362.068967586207 5120.689655862069 -1.0
4362.068967586207 5310.344827931034 -1.0
4362.068967586207 5500.0 -1.0

4551.724139655173 1e-05 670.824626765366
4551.724139655173 189.65518206896553 667.0533921678846
4551.724139655173 379.31035413793103 655.7136269931377
4551.724139655173 568.9655262068966 636.7278395084138
4551.724139655173 758.6206982758621 609.9691541332033
4551.724139655173 948.2758703448276 575.2645959898125
4551.724139655173 1137.9310424137932 532.399440743849
4551.724139655173 1327.5862144827586 481.122411348284
4551.724139655173 1517.2413865517242 421.15147935431196
4551.724139655173 1706.8965586206898 352.1800246975322
4551.724139655173 1896.5517306896552 273.88312268151776
4551.724139655173 2086.206902758621 185.92375592114968
4551.724139655173 2275.8620748275866 87.95878795385943
4551.724139655173 2465.517246896552 -1.0
4551.724139655173 2655.1724189655174 -1.0
4551.724139655173 2844.8275910344832 -1.0
4551.724139655173 3034.4827631034486 -1.0
4551.724139655173 3224.137935172414 -1.0
4551.724139655173 3413.79310724138 -1.0
4551.724139655173 3603.4482793103452 -1.0
4551.724139655173 3793.1034513793106 -1.0
4551.724139655173 3982.7586234482765 -1.0
4551.724139655173 4172.413795517241 -1.0
4551.724139655173 4362.068967586207 -1.0
4551.724139655173 4551.724139655173 -1.0
4551.724139655173 4741.379311724138 -1.0
4551.724139655173 4931.034483793103 -1.0
4551.724139655173 5120.689655862069 -1.0
4551.724139655173 5310.344827931034 -1.0
4551.724139655173 5500.0 -1.0

4741.379311724138 1e-05 481.1224138582006
4741.379311724138 189.65518206896553 477.15087248689656
4741.379311724138 379.31035413793103 465.2133069821266
4741.379311724138 568.9655262068966 445.2414660129175
4741.379311724138 758.6206982758621 417.1234859707947
4741.379311724138 948.2758703448276 380.7066246375821
4741.379311724138 1137.9310424137932 335.80089967643744
4741.379311724138 1327.5862144827586 282.1834646096472
4741.379311724138 1517.2413865517242 219.6035346422418
4741.379311724138 1706.8965586206898 147.7876687731864
4741.379311724138 1896.5517306896552 66.4452234388209
4741.379311724138 2086.206902758621 -1.0
4741.379311724138 2275.8620748275866 -1.0
4741.379311724138 2465.517246896552 -1.0
4741.379311724138 2655.1724189655174 -1.0
4741.379311724138 2844.8275910344832 -1.0
4741.379311724138 3034.4827631034486 -1.0
4741.379311724138 3224.137935172414 -1.0
4741.379311724138 3413.79310724138 -1.0
4741.379311724138 3603.4482793103452 -1.0
4741.379311724138 3793.1034513793106 -1.0
4741.379311724138 3982.7586234482765 -1.0
4741.379311724138 4172.413795517241 -1.0
4741.379311724138 4362.068967586207 -1.0
4741.379311724138 4551.724139655173 -1.0
4741.379311724138 4741.379311724138 -1.0
4741.379311724138 4931.034483793103 -1.0
4741.379311724138 5120.689655862069 -1.0
4741.379311724138 5310.344827931034 -1.0
4741.379311724138 5500.0 -1.0

4931.034483793103 1e-05 273.88312618374266
4931.034483793103 189.65518206896553 269.7278705255426
4931.034483793103 379.31035413793103 257.2418628026676
4931.034483793103 568.9655262068966 236.36485934809934
4931.034483793103 758.6206982758621 206.99803591686486
4931.034483793103 948.2758703448276 169.00628475433996
4931.034483793103 1137.9310424137932 122.221286003552
4931.034483793103 1327.5862144827586 66.44522435190493
4931.034483793103 1517.2413865517242 1.4550045094410962
4931.034483793103 1706.8965586206898 -1.0
4931.034483793103 1896.5517306896552 -1.0
4931.034483793103 2086.206902758621 -1.0
4931.034483793103 2275.8620748275866 -1.0
4931.034483793103 2465.517246896552 -1.0
4931.034483793103 2655.1724189655174 -1.0
4931.034483793103 2844.8275910344832 -1.0
4931.034483793103 3034.4827631034486 -1.0
4931.034483793103 3224.137935172414 -1.0
4931.034483793103 3413.79310724138 -1.0
4931.034483793103 3603.4482793103452 -1.0
4931.034483793103 3793.1034513793106 -1.0
4931.034483793103 3982.7586234482765 -1.0
4931.034483793103 4172.413795517241 -1.0
4931.034483793103 4362.068967586207 -1.0
4931.034483793103 4551.724139655173 -1.0
4931.034483793103 4741.379311724138 -1.0
4931.034483793103 4931.034483793103 -1.0
4931.034483793103 5120.689655862069 -1.0
4931.034483793103 5310.344827931034 -1.0
4931.034483793103 5500.0 -1.0

5120.689655862069 1e-05 49.17997952190717
5120.689655862069 189.65518206896553 44.85616982187321
5120.689655862069 379.31035413793103 31.86685745834302
5120.689655862069 568.9655262068966 10.158796407925365
5120.689655862069 758.6206982758621 -1.0
5120.689655862069 948.2758703448276 -1.0
5120.689655862069 1137.9310424137932 -1.0
5120.689655862069 1327.5862144827586 -1.0
5120.689655862069 1517.2413865517242 -1.0
5120.689655862069 1706.8965586206898 -1.0
5120.689655862069 1896.5517306896552 -1.0
5120.689655862069 2086.206902758621 -1.0
5120.689655862069 2275.8620748275866 -1.0
5120.689655862069 2465.517246896552 -1.0
5120.689655862069 2655.1724189655174 -1.0
5120.689655862069 2844.8275910344832 -1.0
5120.689655862069 3034.4827631034486 -1.0
5120.689655862069 3224.137935172414 -1.0
5120.689655862069 3413.79310724138 -1.0
5120.689655862069 3603.4482793103452 -1.0
5120.689655862069 3793.1034513793106 -1.0
5120.689655862069 3982.7586234482765 -1.0
5120.689655862069 4172.413795517241 -1.0
5120.689655862069 4362.068967586207 -1.0
5120.689655862069 4551.724139655173 -1.0
5120.689655862069 4741.379311724138 -1.0
5120.689655862069 4931.034483793103 -1.0
5120.689655862069 5120.689655862069 -1.0
5120.689655862069 5310.344827931034 -1.0
5120.689655862069 5500.0 -1.0

5310.344827931034 1e-05 -1.0
5310.344827931034 189.65518206896553 -1.0
5310.344827931034 379.31035413793103 -1.0
5310.344827931034 568.9655262068966 -1.0
5310.344827931034 758.6206982758621 -1.0
5310.344827931034 948.2758703448276 -1.0
5310.344827931034 1137.9310424137932 -1.0
5310.344827931034 1327.5862144827586 -1.0
5310.344827931034 1517.2413865517242 -1.0
5310.344827931034 1706.8965586206898 -1.0
5310.344827931034 1896.5517306896552 -1.0
5310.344827931034 2086.206902758621 -1.0
5310.344827931034 2275.8620748275866 -1.0
5310.344827931034 2465.517246896552 -1.0
5310.344827931034 2655.1724189655174 -1.0
5310.344827931034 2844.8275910344832 -1.0
5310.344827931034 3034.4827631034486 -1.0
5310.344827931034 3224.137935172414 -1.0
5310.344827931034 3413.79310724138 -1.0
5310.344827931034 3603.4482793103452 -1.0
5310.344827931034 3793.1034513793106 -1.0
5310.344827931034 3982.7586234482765 -1.0
5310.344827931034 4172.413795517241 -1.0
5310.344827931034 4362.068967586207 -1.0
5310.344827931034 4551.724139655173 -1.0
5310.344827931034 4741.379311724138 -1.0
5310.344827931034 4931.034483793103 -1.0
5310.344827931034 5120.689655862069 -1.0
5310.344827931034 5310.344827931034 -1.0
5310.344827931034 5500.0 -1.0

5500.0 1e-05 -1.0
5500.0 189.65518206896553 -1.0
5500.0 379.31035413793103 -1.0
5500.0 568.9655262068966 -1.0
5500.0 758.6206982758621 -1.0
5500.0 948.2758703448276 -1.0
5500.0 1137.9310424137932 -1.0
5500.0 1327.5862144827586 -1.0
5500.0 1517.2413865517242 -1.0
5500.0 1706.8965586206898 -1.0
5500.0 1896.5517306896552 -1.0
5500.0 2086.206902758621 -1.0
5500.0 2275.8620748275866 -1.0
5500.0 2465.517246896552 -1.0
5500.0 2655.1724189655174 -1.0
5500.0 2844.8275910344832 -1.0
5500.0 3034.4827631034486 -1.0
5500.0 3224.137935172414 -1.0
5500.0 3413.79310724138 -1.0
5500.0 3603.4482793103452 -1.0
5500.0 3793.1034513793106 -1.0
5500.0 3982.7586234482765 -1.0
5500.0 4172.413795517241 -1.0
5500.0 4362.068967586207 -1.0
5500.0 4551.724139655173 -1.0
5500.0 4741.379311724138 -1.0
5500.0 4931.034483793103 -1.0
5500.0 5120.689655862069 -1.0
5500.0 5310.344827931034 -1.0
5500.0 5500.0 -1.0
    };
\end{axis}
\end{tikzpicture}